\documentclass[12pt]{iopart}
\usepackage{lscape}
\usepackage{iopams}
\usepackage{graphicx}
\usepackage{epstopdf}
\usepackage{color}
\usepackage[usenames,dvipsnames]{xcolor}

\begin{document}
\title[A PCFI approach for describing electron correlation in atoms]{A Partitioned Correlation Function Interaction approach for describing electron correlation in atoms}
\author{S. Verdebout$^{1}$, P. Rynkun$^{2}$, P. J\"onsson$^{3}$, G. Gaigalas$^{2,4}$, C.~Froese Fischer$^{5}$ and M. Godefroid$^{1}$}

\address{$^{1}$ Chimie Quantique et Photophysique, CP160/09, Universit\'e Libre de Bruxelles, \\ 
Av. F.D. Roosevelt 50, B-1050 Brussels, Belgium} 

\address{$^{2}$ Vilnius Pedagogical University, Student\c{u} 39, Vilnius, LT-08106, Lithuania}

\address{$^{3}$  Group for Materials Science and Applied Mathematics, Malm\"{o} University, 20506 Malm\"{o}, Sweden }

\address{$^{4}$ Vilnius University Research Institute of Theoretical Physics and Astronomy, \\ 
A. Go\v{s}tauto 12, LT-01108 Vilnius, Lithuania}

\address{$^{5}$ Department of Computer Science, Vanderbilt University, Nashville, TN 37235, USA }

\ead{mrgodef@ulb.ac.be}
\begin{abstract}
The traditional multiconfiguration Hartree-Fock (MCHF) and configuration interaction (CI) methods are based on a single orthonormal orbital basis.
For atoms with many closed core shells, or complicated shell structures, a large orbital basis is needed to saturate the different electron correlation effects such as valence, core-valence correlation and correlation within the core shells.  
The large orbital basis leads to massive configuration state function (CSF) expansions that are difficult to handle, even on large computer systems.   
We show that it is possible to relax the orthonormality restriction on the orbital basis and break down the originally very large calculations to a series of smaller calculations
that can be run in parallel. Each calculation determines a partitioned correlation function (PCF) that accounts for a specific correlation effect. 
The PCFs are built on optimally localized orbital sets and are added to a zero-order multireference (MR) function to form a total wave function. The expansion coefficients of the PCFs are determined from a
low dimensional generalized eigenvalue problem. The interaction and overlap matrices  are computed using a biorthonormal
transformation technique (Verdebout~{\it et al.}, J. Phys. B: At. Mol. Phys. {\bf 43} (2010) 074017). The new method, called partitioned correlation function interaction (PCFI), converges rapidly with respect to the orbital basis and gives total energies that
are lower than the ones from ordinary MCHF and CI calculations.
The PCFI method is also very flexible when it comes to targeting different electron correlation effects. Focusing our attention on neutral lithium, we show that by dedicating a PCF to the single 
excitations from the core, spin- and orbital-polarization effects can be captured very efficiently, leading to highly improved convergence patterns for hyperfine parameters compared with MCHF calculations based on a single orthogonal radial orbital basis.

By collecting separately optimized PCFs  to correct the MR function, the variational degrees of freedom in the relative mixing coefficients of the CSFs building the PCFs are inhibited. The
constraints on the mixing coefficients lead to small off-sets in computed properties such as hyperfine structure, isotope shift and transition rates, with respect to the correct values.   
By (partially) deconstraining the mixing coefficients one converges to the correct limits and keeps the tremendous advantage of improved convergence rates that comes from the use of several orbital sets. Reducing ultimately each PCF to a single CSF with its own orbital basis leads to a non-orthogonal configuration interaction approach. Various perspectives of the new method are given.
\end{abstract}

\submitto{\jpb}

\vfill
\noindent{\hfill \bf \today}

\maketitle

\section{Introduction}

The electron correlation energy of an atom has been defined by L\"owdin~\cite{Low:55b} as the difference 
between the exact nonrelativistic energy eigenvalue of the electronic Schr\"odinger equation and the energy of the single configuration state function (CSF) approximation, commonly called the Hartree-Fock energy. 
In line with this definition we think of electron correlation effects as those manifesting themselves beyond the Hartree-Fock  approximation. It is  useful to subdivide further and consider both static (nondynamical) and dynamical correlation~\cite{Moketal:96a}. Static correlation arises from near-degeneracies of the Hartree-Fock occupied and unoccupied orbitals. Systems with significant static correlation are poorly described by a single CSF and are said to have a strong multireference character. Dynamical correlation is due to the reduction in the repulsion energy related to the electron-electron cusp and is a short-range effect.

Accurate description of electron correlation remains a major challenge in atomic structure calculations. To meet this challenge a number of different methods have been developed such as many-body perturbation theory (MBPT) \cite{Joh:2007a,Lin:2011a}, combinations of configuration interaction and many-body perturbation (CI+MBPT) \cite{DzuFla:07a,Johetal:2008a,Safetal:09a}, and coupled cluster (CC)~\cite{Knoetal:2000a,Bar:2010a,Dasetal:2012a} theories. Different kinds of variational methods have also been used, and one may specially note Hylleraas-type calculations, that explicitly include the interelectron distance $r_{12}$ in the construction of the wave function~\cite{Hyl:64a,SimHag:2009a,Kometal:95a,Kinetal:2011a}. In quantum chemistry, variational complete active space self-consistent field (CASSCF) methods are quite successful for describing small and medium-size molecules, but are not sufficient when dynamical correlation must be included~\cite{Roo:87a}. The latter are treated through second-order perturbation theory using a single or multireference state as the zero-order approximation. Combined variational multireference and second order M\"oller-Plesset perturbation calculations have also been applied very successfully by Ishikawa and co-workers~\cite{Viletal:08a,Ishetal:2010a} to obtain accurate transition energies for a number of atomic systems. In this paper we will critically examine variational multiconfiguration methods, such as multiconfiguration Hartree-Fock (MCHF) combined with configuration interaction (CI).

Multiconfiguration methods are quite general, and can be directly applied to excited and open-shell structures across the whole periodic table. By including the most important closely degenerate CSFs to form a multireference (MR) expansion, the static correlation is efficiently captured. Dynamical correlation is accounted for by 
adding, to the MR expansion, CSFs obtained by single (S) and double (D) excitations from the CSFs in the MR to an increasing set of active orbitals. The CSFs generated in this way build a space that we refer to as the correlation function (CF) space and it is convenient to think of multiconfiguration expansions as something built from CSFs in the MR space and in the CF space.
Due to restrictions in the Racah or the Slater determinant algebra underlying the construction of the energy expressions, the orbitals are usually required to be orthonormal. 
Such an orthonormal orbital basis is not very efficient for larger systems. 
Let us consider an atomic system, for example Ca, with several closed shells.  
To describe the dynamic correlation in the $1s$ shell we tailor an orbital set for which some of the orbitals should have a large overlap with the $1s$ radial orbital.
Due to the orthogonality restrictions of the orbitals, the correlation in the $2s$ shell needs to be described in terms of the previous radial orbitals, tailored for describing correlation within the $1s$ shell, as well
as some new radial orbitals that are overlapping with the $2s$ orbital etc.  To capture the dynamic correlation between electrons in all the different shells, the orbital basis needs to be extended to a large number of orbitals for each symmetry, leading to massive CSF expansions~\cite{Caretal:2010a}. 
 This is in effect a scaling wall~\cite{Weietal:2007a} that has been difficult to get around. 
In practice the electrons in the atom are considered as either core or outer valence electrons resulting in valence-valence, core-valence, and core-core types of SD excitations with the latter often neglected.
 Another general problem with variational methods is that they are entirely based on the energy functional, and properties not strongly coupled to this functional may be inadequately described by the resulting wave function. As an example we  consider the hyperfine interaction. The CSFs that are responsible for the important spin- and orbital-polarization effects are relatively unimportant for the total energy, and thus the orbital basis from the variational calculation may be spatially localized in such a way that the above effects are not captured. Alternatively, a very large orbital basis is needed to achieve convergence for these properties, leading to CSF expansions that grow unmanageably large. 

The present work is an extension of a previous study of correlation energy in beryllium~\cite{Veretal:2010a}. Based on a fast transformation technique, originally proposed by Malmqvist and collaborators~\cite{Mal:86a,Olsetal:95a}, we show that it is possible to relax the orthonormality restriction on the orbital basis and use several 
mutually non-orthogonal orbital basis sets that are better adapted to the short range nature of the dynamical correlation. The gained freedom also makes it possible to tailor an orbital basis for capturing effects weakly connected to energy, improving convergence properties of atomic
properties other than the energy. 
Partitioning the CF space into several subspaces, and using different orbital sets optimized for the different partitions, may be one way around the scaling wall associated with single orthonormal orbital sets.

\section{Partitioning the MCHF problem}
Starting from the non-relativistic Hamiltonian for an $N$-electron system
\begin{equation}
H = \sum_{i=1}^N \left[ - \frac{1}{2} \nabla_i^2 - \frac{Z}{r_i} \right] + \sum_{i<j}^N \frac{1}{r_{ij}} \; ,
\end{equation}
the multiconfiguration Hartree-Fock (MCHF) approach determines an approximate wave function of the form
\begin{equation} \label{MCHF_exp}
\Psi(\gamma\ LS^\pi) = \sum_{i=1}^M  \; c_i~\Phi_i(\gamma_i\ LS^\pi) \; ,
\end{equation}
in which a CSF, $\Phi_i(\gamma_i\ LS^\pi)$, belongs  either to the MR space or to the CF space. All CSFs have a given parity ($\pi$) and $LS$ symmetry and they are built from a common basis of one-electron spin-orbitals
\begin{equation}
\phi(nlm_lm_s) = \frac{1}{r} P(nl\,;\,r) Y_{lm_l}(\theta, \varphi)\chi_{m_s},
\end{equation}
where the radial functions $P(nl\,;\,r)$ are to be determined \cite{Froetal:97b}. 
For the approximate wave function~\eref{MCHF_exp}, the integro-differential MCHF equations have the form
\begin{eqnarray} \label{MCHF_eq}
\left\{ \frac{d^2}{dr^2} + \frac{2}{r} \left[Z - Y(nl\,;\,r)\right]  - 
\frac{l(l+1)}{r^2} - \epsilon_{nl,nl} \right\} P(nl\,;\,r) \nonumber \\
                    = \frac{2}{r} X(nl\,;\,r) + \sum_{n'\ne n} \epsilon_{nl,n'l} P(n'l\,;\,r)
\end{eqnarray}
for the unknown radial functions \cite{Froetal:97b}. The equations are coupled to each other through the direct $Y$ and exchange $X$ potentials and the Lagrange multipliers $\epsilon_{nl,n'l}$. The Lagrange multipliers force the radial orbitals to be orthonormal within the same $l$ subspace. Under these conditions the configuration state functions are orthonormal 
\begin{equation} 
\langle \Phi_i| \Phi_j \rangle = \delta_{i,j}.
\end{equation}
The mixing coefficients appearing in the expansion over CSFs also enter in the form of the potentials and are determined by solving the configuration interaction (CI) problem 
\begin{equation}
{\bf Hc} = E {\bf c},
\label{CI}
\end{equation}
with $H_{ij} = \langle \Phi_i|H| \Phi_j \rangle$ being the Hamiltonian matrix and ${\bf c} = (c_1,c_2,\ldots,c_M)^t$ the column vector of mixing coefficients. For a given set of mixing coefficients, the equations~\eref{MCHF_eq} are solved by the self-consistent field (SCF) procedure. The SCF and CI problems are solved, one after the other, until convergence of both the radial functions and the selected CI-eigenvector is achieved.  
 
 The strong coupling between the CSF expansion and the resulting optimized orbital basis is well known~\cite{Godetal:98a}. In the variational multiconfiguration approach indeed, the orbitals adapt spatially to account for the specific correlation effect targeted by the tailored expansion. 
 In the present work, we investigate the possibility of breaking down the computational task into subtasks
 by partitioning the CF space into different subspaces, each targeting a specific correlation effect such as valence-correlation, core-valence or correlation within shells in the core, and performing separate MCHF calculations for each expansion built on the MR space and a CF subspace.
In a final step the wave function is expanded in a basis consisting of CSFs from the MR space and functions built in each of the CV subspaces.
The expansion coefficients are obtained by computing the Hamiltonian and overlap matrices and solving the corresponding generalized eigenvalue problem.
The computation of the matrix elements between functions in the different subspaces, and this is the crucial point, is made possible by the biorthogonal transformation \cite{Mal:86a,Olsetal:95a}.

The above  scheme  offers various advantages, resembling the ``Divide and Conquer'' strategy 
-~i)~from the computational point of view, smaller subtasks can be run in parallel, -~ii)~the resulting orbital basis sets are better adapted for capturing efficiently electron correlation, with the hope of getting a satisfactory accuracy for the desired property before reaching the scaling wall, -~iii)~the coupling between the subspaces reduces to a reasonably small dimension eigenvalue problem.  

 \section{The PCFI approach}
\label{PCFI}
The efficiency of the method with a partitioned CF space was shown in our first paper~\cite{Veretal:2010a}, when targeting the total energy of the ground state of neutral beryllium.  A multi-reference (MR) expansion 
\begin{eqnarray} 
| \Psi^{\textsc{mr}} (\gamma\ LS^\pi) \rangle = \sum_{j=1}^m a_j~| \Phi_j^{\textsc{mr}}(\gamma_j\ LS^\pi) \rangle \; ,
 \label{eq:MR}
\end{eqnarray}
limited to the major contributions to valence correlation, including the near-degenerate $1s^22s^2$ and $1s^22p^2$ configurations of the  Layzer's complex, was corrected by three Pair Correlation Functions $\vert \Lambda_\textsc{vv} \rangle$, $\vert \Lambda_\textsc{cv} \rangle$, $\vert \Lambda_\textsc{cc} \rangle$ built by allowing single and double excitations from specific subshells of the MR configuration state functions to  a given orbital active set (AS),  and  specifically tailored to describe the valence (VV), core-valence (CV) and core (CC) correlation effects. The final wave function  
\begin{equation}
 | \Psi(\gamma\ LS^\pi) \rangle = | \Psi^{\textsc{mr}} (\gamma\ LS^\pi) \rangle
 +  \alpha_\textsc{vv} \;  \vert \Lambda_\textsc{vv} \rangle 
 +  \alpha_\textsc{cv} \; \vert \Lambda_\textsc{cv} \rangle 
 +  \alpha_\textsc{cc} \; \vert \Lambda_\textsc{cc} \rangle 
 \label{eq:orig}
\end{equation}
yielded a lower energy than the traditional MCHF method based on a very large CSF expansion.  
\\

In the present work, we generalize the approach by introducing the ``{\it Partitioned} Correlation Functions'' instead of ``{\it Pair} Correlation Functions'', preserving the PCF acronym, but allowing more flexible building rules for each PCF.
 For describing unambiguously the CSFs content of a given PCF, we first define a ``pure'' PCF as a CSFs expansion containing  only one kind of excitations, ie. single (S), {\it or} double~(D), {\it or} triple (T),\ldots excitations. 
We write such a Partitioned Correlation Function   $\Lambda_{\{i\}}$, where the subscript $\{i\}$ specifies the set of occupied shells that are excited to a given active set. For instance, a PCF including only double excitations from the $n_1l_1$ and $n_2l_2$ subshells\footnote{Note that $n_1l_1$ and $ n_2l_2$  may refer to equivalent or non-equivalent electrons.} is written $\Lambda_{n_1l_1 \; n_2l_2}$. A PCF could also be ``hybrid'' if containing different kinds of excitations. Such a PCF may be seen as a superposition of ``pure'' PCFs and is written  as $\Lambda_{\{i\}-\{j\}-\ldots}$, where for each family of excitations $\{i\}, \{j\}, \ldots$,  the letters sequence appearing in the subscript specify the $nl$-labels of the MR electrons that undergo the excitations.  
It is clear that a Partitioned Correlation Function in our approach does not fit with the usual definition of a pair correlation function~\cite{Fro:77a,LinSal:80a}.

According to this notation, 
a PCF representing  single excitations from $n_1l_1$ and double excitations from the $n_1l_1$ and $n_2l_2$ subshells, is written as
\begin{equation}
\vert \Lambda_{n_1l_1-n_1l_1n_2l_2} \rangle = \sum_{nl} \alpha_{nl} | \Phi_{n_1l_1}^{nl} \rangle + \sum_{nl,n'l'} \alpha_{nl,n'l'} | \Phi_{n_1l_1,n_2l_2}^{nl,n'l'} \rangle 
\label{eq:ex_not_1}
\end{equation}
where the first summation corresponds to all possible single excitations from the $n_1l_1$ shell for each CSF belonging to the MR and the second one, to all possible double excitations from the $n_1l_1$ and $n_2l_2$ shells for each CSF belonging to the MR.
One can then rewrite the beryllium ground state wave function \eref{eq:orig} as
\[
 | \Psi(\gamma\ LS^\pi) \rangle = | \Psi^{\textsc{mr}} (\gamma\ LS^\pi) \rangle  \]
\[ \hspace*{2cm} +   \alpha_\textsc{vv} \; \vert \Lambda_{v-vv} \rangle 
 +  \alpha_\textsc{cv} \; \vert \Lambda_{1s \; v} \rangle 
 +  \alpha_\textsc{cc} \; \vert \Lambda_{1s-1s 1s} \rangle 
\]
where $v$ stands for any occupied valence subshell of the MR set. 
In order to avoid having the same CSF in two different partitioned correlation functions, we consider the partitioning of the configuration space into disjoint sets. Another possible partition satisfying this property could be
\[
 | \Psi(\gamma\ LS^\pi) \rangle = | \Psi^{\textsc{mr}} (\gamma\ LS^\pi) \rangle  \]
\[ \hspace*{2cm} +   \alpha'_\textsc{vv} \; \vert \Lambda_{v-vv} \rangle 
 +  \alpha' _\textsc{cv} \; \vert \Lambda_{1s - 1s \; v} \rangle 
 +  \alpha'_\textsc{cc} \; \vert \Lambda_{1s 1s} \rangle 
\]
where the single excitations $(1s \rightarrow n_1 l_1)$ have arbitrarily moved from the CC to the CV correlation function subspace.
In the most general case, the  MR function \eref{eq:MR} is corrected by $p$ PCFs
\begin{equation}
 | \Psi(\gamma\ LS^\pi) \rangle = | \Psi^{\textsc{mr}} (\gamma\ LS^\pi) \rangle
 + \sum_{i=1}^p \alpha_i \;  | \Lambda_i \rangle \; ,
 \label{eq:formal_PCF_i}
\end{equation}
each of the PCFs corresponding to a given partition of the CF space:
\begin{equation}
 | \Lambda \rangle  = \sum_j^{dim(\Lambda)} c_j^\Lambda \; \vert \Phi_j^\Lambda \rangle \; .
 \label{Lambda}
\end{equation}
We use the notation $\Psi^{\Lambda}$ for the function consisting of the MR function and one of the correcting PCFs~$\Lambda$
\begin{equation}
 | \Psi^{\Lambda}(\gamma\ LS^\pi) \rangle = \sum_{j=1}^m a^\Lambda_j~| \Phi_i^{\textsc{mr}}(\gamma_j\ LS^\pi) \rangle
 + \sum_j^{dim(\Lambda)} c_j^\Lambda \;  | \Phi_j^{\Lambda} \rangle \; .
 \label{eq:MR-PCF}
\end{equation}
In our approach, this function is obtained by solving the corresponding  MCHF equations~\eref{MCHF_eq}-\eref{CI} to optimize the $\Lambda$-PCF orbital set and mixing coefficients.
Such a calculation that optimizes the MR eigenvector composition $\{ a^{\Lambda}_j \}$ with orbitals frozen to the MR-MCHF solution~\eref{eq:MR}, the mixing coefficients $\{ c^{\Lambda}_j \}$ and the $\Lambda$-PCF radial functions, will be referred as a MR-PCF calculation.
As far as the notation is concerned, we will underline when necessary the orbitals that are kept frozen during the self-consistent-field process.
Solving the MCHF problem~\eref{eq:MR-PCF} for each $\Lambda_i$~$(i=1,2,\ldots,p)$ produces $p$ mutually non-orthonormal one-electron orbital sets.
Each of the orbital sets will be optimally localized for the correlation effect described by the corresponding PCF expansion. 
Assuming the CSFs of the MR and CF spaces orthonormal and 
$\langle \Psi^{\Lambda} \vert \Psi^{\Lambda} \rangle =1$,  we have
\[
\langle \Psi^{\Lambda} \vert \Psi^{\Lambda} \rangle =  \langle  \Psi^{\textsc{mr}} \vert \Psi^{\textsc{mr}} \rangle +  \langle \Lambda \vert \Lambda \rangle  = \sum_{j=1}^m  (a^\Lambda_j)^2 +  \sum_j^{dim(\Lambda) } (c_j^\Lambda)^2 =  1 \; ,
\]
revealing that $\langle \Lambda \vert \Lambda \rangle \neq 1$. To keep a natural interpretation of the PCF weights, we renormalize each PCF according to
\[
\overline{\Lambda} = \frac{1}{\sqrt{\sum_j (c_j^\Lambda)^2}} \; \Lambda \; .
\]

The PCFI approach consists in regrouping the $m$ components of the MR space and the $p$ CF subspaces in an {\it a priori} low-dimension interaction matrix to get a compact representation of the total wave function
\begin{eqnarray}
| \Psi \rangle = \sum_{i=1}^m a_i  | \Phi^{\textsc{mr}}_i \rangle + \sum^p_{j} b_j |\overline{\Lambda}_j \rangle \;,
\label{eq;wfn}
\end{eqnarray}
where the mixing coefficients $\{ a_i \} $ and $\{ b_j \}$ are obtained by solving the generalized eigenvalue problem of dimension $(M=m+p)$ 
\begin{equation}
{\bf Hc} = E{\bf Sc} \; .
\label{GEP}
\end{equation}
The corresponding Hamiltonian matrix may be explicitly written as 
\vspace*{0.2cm}
\begin{equation}
\hspace*{-2cm}
{\bf H} = \left(
\begin{array}{c c c c}
\framebox(190,60)[]{$
\begin{array}{c c c}
\langle  \Phi^{\textsc{mr}}_1 | H |  \Phi^{\textsc{mr}}_1 \rangle & \cdots & \langle  \Phi^{\textsc{mr}}_1 | H |  \Phi^{\textsc{mr}}_m \rangle \\
\vdots & \ddots & \vdots \\
\langle  \Phi^{\textsc{mr}}_m | H |  \Phi^{\textsc{mr}}_1 \rangle & \cdots & \langle  \Phi^{\textsc{mr}}_m | H |  \Phi^{\textsc{mr}}_m \rangle \\
\end{array}$} & 
\framebox(70,60)[]{$
\begin{array}{c}
\langle  \Phi^{\textsc{mr}}_1 | H | \overline{\Lambda}_1 \rangle \\
\vdots \\
\langle  \Phi^{\textsc{mr}}_m | H | \overline{\Lambda}_1 \rangle \\
\end{array}$} & \cdots & 
\framebox(70,60)[]{$
\begin{array}{c}
\langle  \Phi^{\textsc{mr}}_1 | H | \overline{\Lambda}_p \rangle \\
\vdots \\
\langle  \Phi^{\textsc{mr}}_m | H | \overline{\Lambda}_p \rangle \\
\end{array}$}\\
\vdots & \framebox(70,30)[]{$ \langle \overline{\Lambda}_1 | H | \overline{\Lambda}_1 \rangle $}  & \cdots & \dashbox{1.0}(70,30)[]{$ \langle \overline{\Lambda}_1 | H | \overline{\Lambda}_p \rangle $} \\  
\vdots & \vdots & \ddots & \vdots \\
\vdots & \cdots  & \cdots & \framebox(70,30)[]{$ \langle \overline{\Lambda}_p | H | \overline{\Lambda}_p \rangle $} \\
\end{array}
\right)
\label{eq:big_mat}
\end{equation}
The matrix dimension $M$ is simply given by the sum of the number of CSFs belonging to the MR ($m$) and the number of PCF functions~($p$).
The overlap matrix has the same structure, with a value of unity on the diagonal thanks to the renormalization ($| \Lambda_{i} \rangle \rightarrow | \overline{\Lambda}_{i} \rangle$). It can be obtained by merely substituting the $H$ operator, appearing in each matrix element, by the unit operator $\hat{1}$. It  reduces to the unit matrix for specific PCF-building rules as shown in Appendix~1.
All plain-line blocks  involve orthonormal orbitals and the construction of the matrix elements between the CSFs in the blocks is based on fast angular integration methods 
developed by Gaigalas {\sl et al.}~\cite{GaiRud:96a,Gaietal:97a}. This holds not only for the diagonal blocks, but also for blocks coupling the CF and the MR spaces since  we do not allow the MR orbitals to vary in the MR-PCF MCHF calculations of~\eref{eq:MR-PCF}. The building of all other blocks of the CF space, surrounded by a dashed line, involve non-orthogonal orbitals arising from independent MCHF calculations and requires therefore the use of biorthonormal transformations before the traditional methods for angular integration  can be applied (see section~6 of~\cite{Veretal:2010a}). By solving this compact eigenvalue problem, we showed in~\cite{Veretal:2010a} that accurate total energies can be obtained. 

However, as we already mentioned in that work, some variational freedom in the coefficients is lost  by the fact that solving \eref{GEP} does not allow relaxation in the relative weights within each PCF. The latter are indeed fixed linear combinations,
\begin{equation}
\vert \overline{\Lambda} \rangle = \sum_k \overline{c}^\Lambda_k \; \vert \Phi_k \rangle
\label{eq:Lambda_frozen}
\end{equation}
 and as such, we will refer to
$\vert \overline{\Lambda} \rangle$ as a {\it constrained} CSFs expansion in the sense that the orbitals and the expansion coefficients are not allowed to change. The coefficients $\{ \overline{c}^\Lambda_k \}$  will be called the {\it constrained coefficients}. The effect associated with this loss of flexibility on a property, that we are investigating in the next section, will be qualified as the {\it constraint effect}. 


\section{The constraint effect }
\label{constraint}

\begin{figure}
  \begin{minipage}[c]{.46\linewidth}
\include{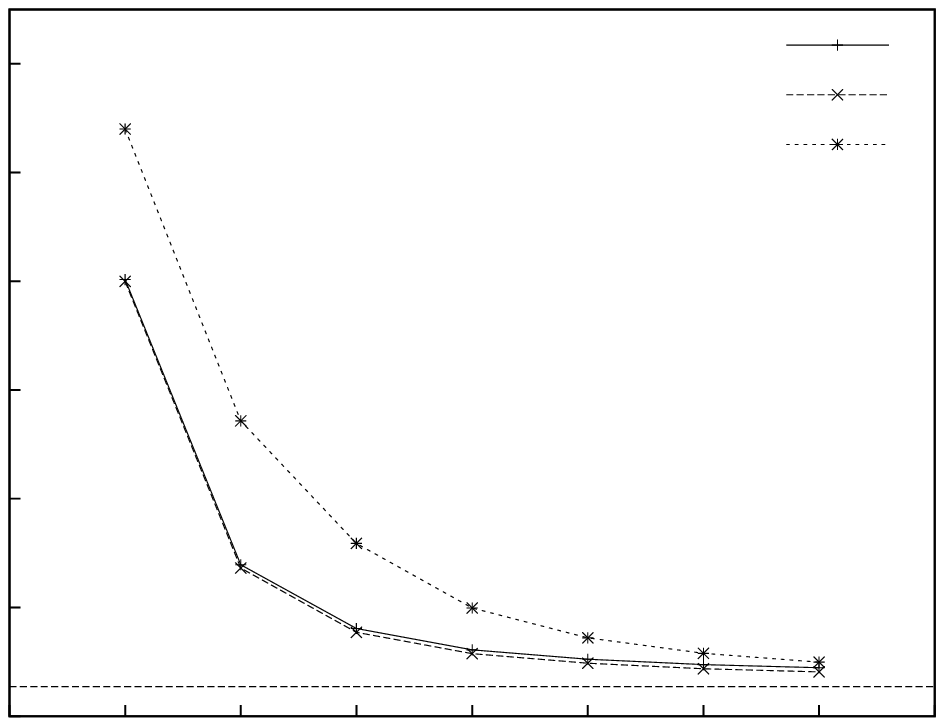}
   \end{minipage} \hfill
   \begin{minipage}[c]{.46\linewidth}
\include{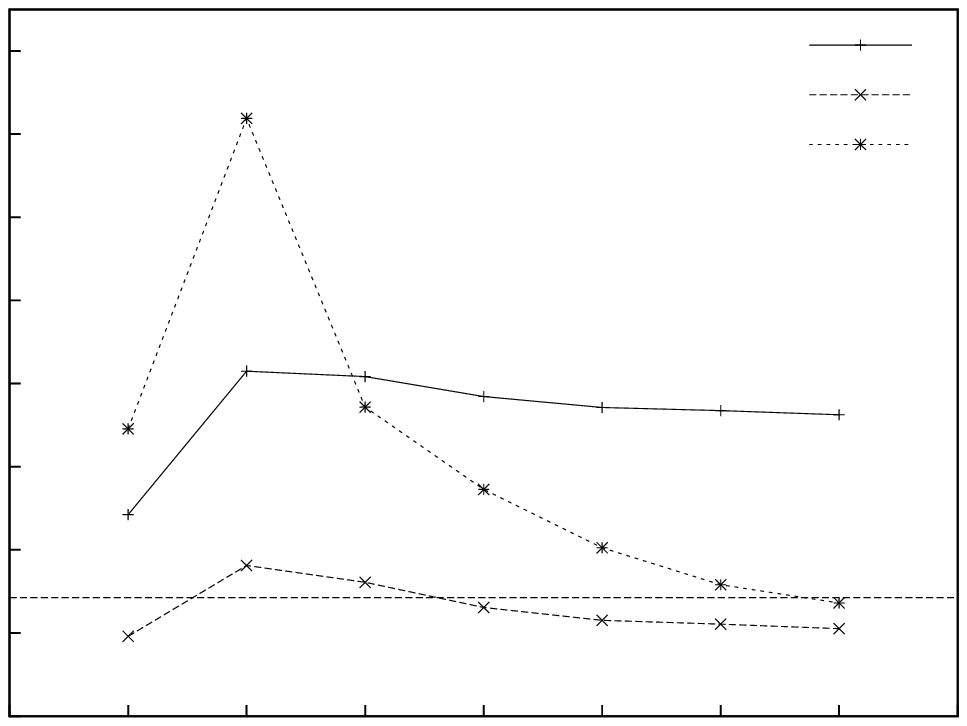}
   \end{minipage}
\caption{Be second excited state. Comparison of the convergence trends between the SD-MR-PCFI, SD-MR-MCHF and SD-MR-DPCFI (see section~\ref{constraint}) approaches for the total energy (left figure) and the SMS parameter (right figure). The horizontal lines correspond to the results of Komasa and Rychlewski~\cite{KomRyc:2001a}.}
\label{fig:Be_1P}
\end{figure}

We were expecting that this {\it constraint} would have a minor impact on all expectation values but extending our previous work~\cite{Veretal:2010a} to $1s^22s2p~^1P^o$ of Be brought a surprise for spectroscopic properties other than the total energy. For this state, we optimized the following MR ($m=6$) function
\[ \mbox{MR} = 1s^2\{2s2p,2s3p,2p3s,2p3d,3s3p,3p3d \; \;  \; ^1P^o \} \;, \]
that we corrected by three PCFs functions ($p=3$) targeting respectively valence, core-valence and core correlation and built on single- and double-excitations (SD) from the multi-reference set
\[
 | \Psi(1s^22s2p~^1P^o) \rangle = | \Psi^{\textsc{mr}} (^1P^o) \rangle  \]
\[ \hspace*{2cm} +   \alpha_\textsc{vv} \; \vert \Lambda_{v-vv} \rangle 
 +  \alpha_\textsc{cv} \; \vert \Lambda_{1s \; v} \rangle 
 +  \alpha_\textsc{cc} \; \vert \Lambda_{1s-1s 1s} \rangle,
\]
where $v$ stands for any orbital of the valence MR set, ie. $v = \{2s,2p,3s,3p,3d \}$.
After solving the MR-PCF problem~\eref{eq:MR-PCF} for the three $\Lambda$'s, we solved the $(M=m+p=9)$ eigenvalue PCFI problem.  We will use the acronyms SD-MR-PCFI and SD-MR-MCHF, respectively, for the present  Partitioned Correlation Function Interaction and for the conventional multi-reference MCHF calculations. 
\Fref{fig:Be_1P} shows the striking difference in behavior between the resulting energy and specific mass shift (SMS) parameter. While the total energies obtained with the SD-MR-PCFI and the SD-MR-MCHF models converge to the same limit, the two approaches give rise to an unexpected off-set on the SMS parameter. 

\begin{figure}[!h]
\begin{center}
\includegraphics[scale=0.5]{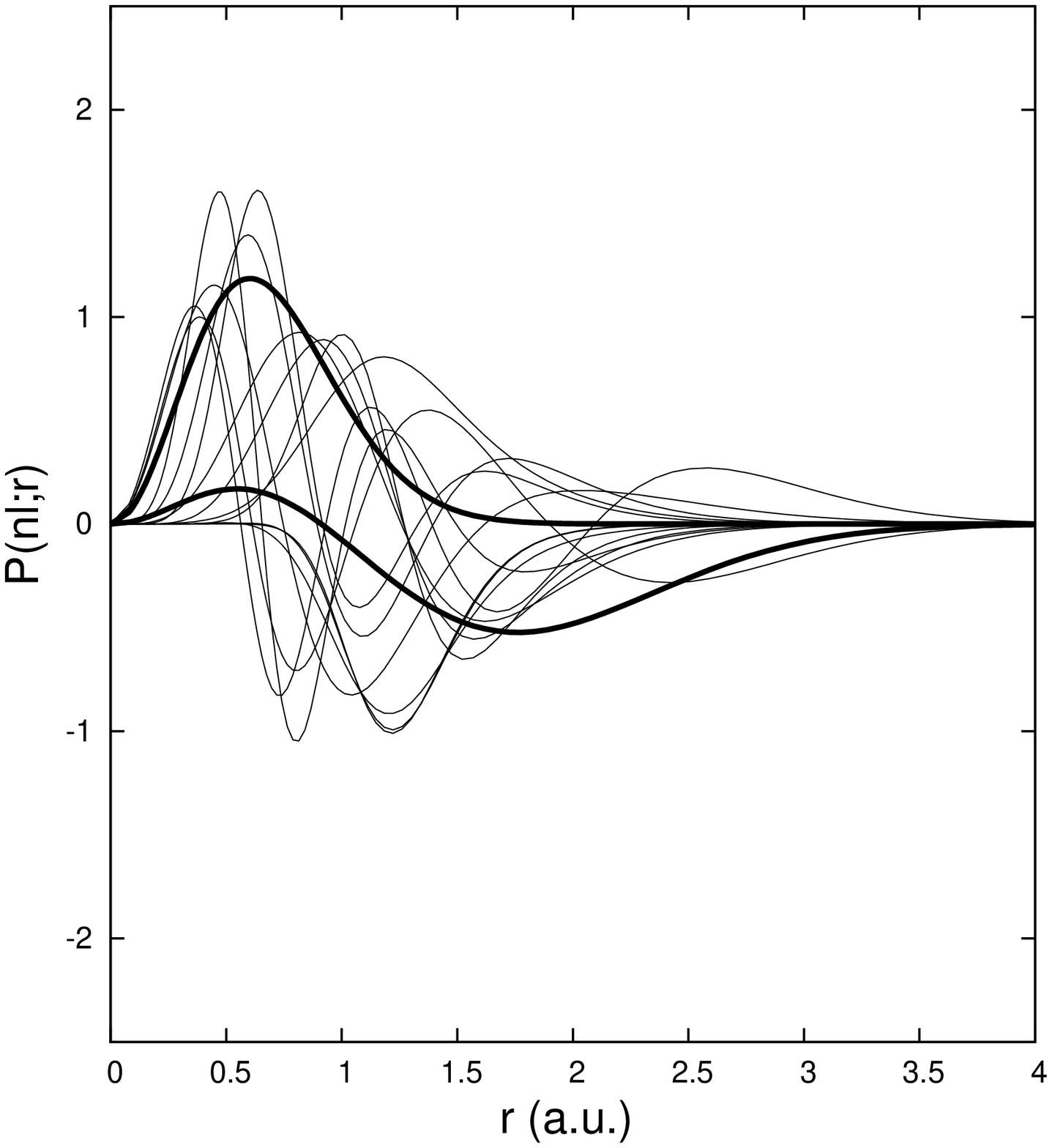}
\\
\includegraphics[scale=0.5]{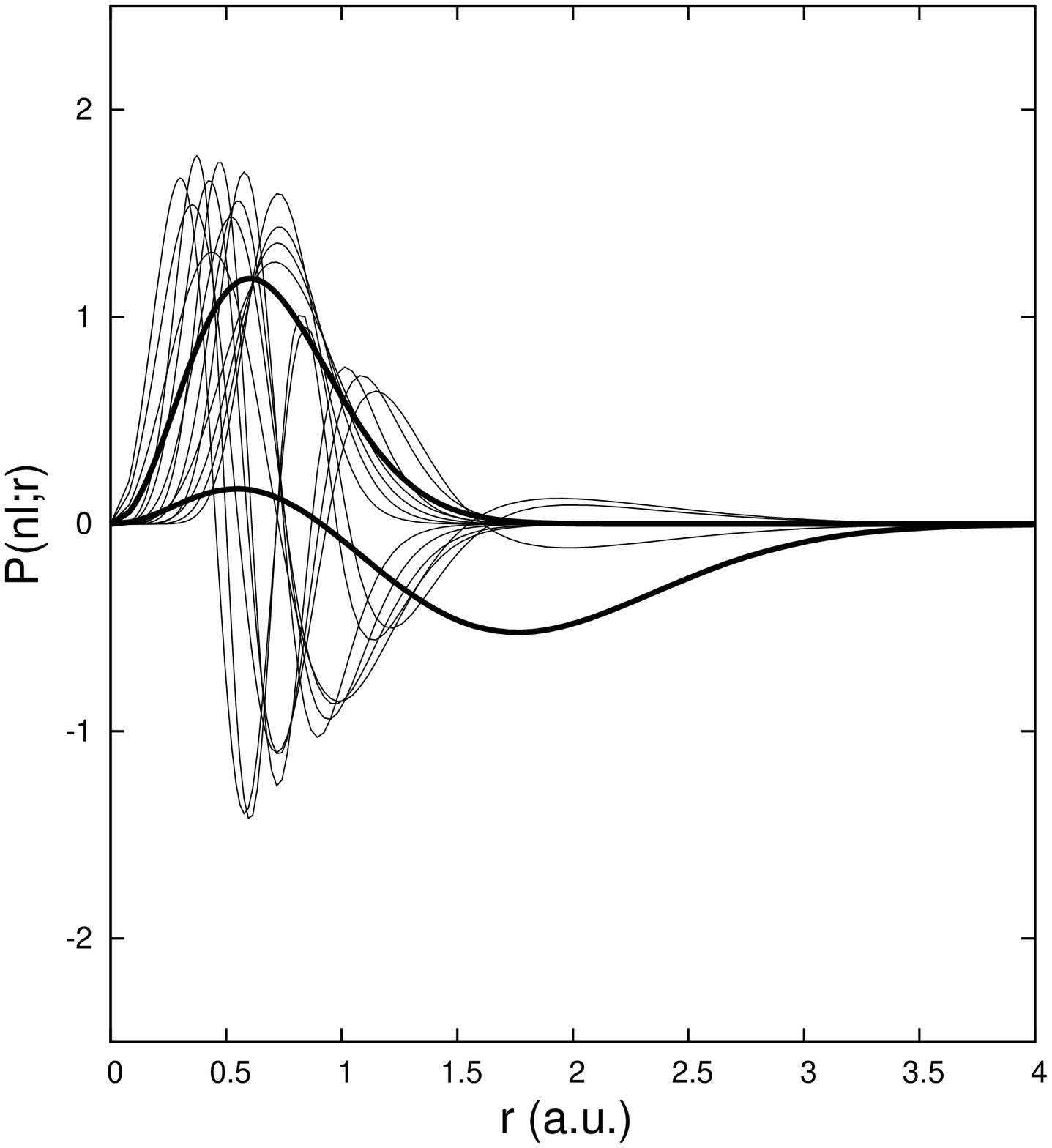}
\caption{The two thick lines correspond to the spectroscopic Hartree Fock $\underline{1s}$ (no node) and $\underline{2s}$ (one node) orbitals of Li $1s^2\,2s~^2S$. The other lines represents the radial functions of the correlation orbitals of the $n=5$ active set. The comparison between the top (core-valence MCHF) and bottom (core-core MCHF) figures illustrates the contraction of the  first few correlation orbitals when going from $\Lambda_{2s-1s2s}$ to $\Lambda_{1s-1s1s}$ (mono)-PCF calculations. }
\label{fig:PCF_orbitals}
\end{center}
\end{figure}

For a deeper understanding of the ins and outs of the PCFI approach and fully appreciating its advantages, we moved to a smaller system: neutral lithium  and its spectroscopic properties. For the lithium ground state, the Hartree-Fock approximation is rather good and the monoreference ($m=1$) $1s^2 2s \; ^2S$ can be taken. We apply the PCFI method using two $(p=2)$ PCFs:~-i) the first one targeting single and double excitations from the core ($1s$) orbital and denoted $\Lambda_{1s-1s1s}$~ , - ii) a second one, $\Lambda_{2s-1s2s}$, targeting single excitations from the $2s$ valence shell and double excitations from the core ($1s$) and valence ($2s$) orbitals. The size of the PCFI matrix is small $(M=3)$.
\Fref{fig:PCF_orbitals} illustrates the MCHF radial functions of the ($n=5$) active set resulting from the two MR-PCF equations applied to $\Lambda_{1s-1s1s}$ and $\Lambda_{2s-1s2s}$. As it was observed in~\cite{Veretal:2010a,Godetal:98a} for other systems, one can realize from this figure that a given PCF orbitals set specifically favors the region of the space occupied in the reference by the electrons undergoing excitations.


\begin{figure}[!ht]
\begin{center}
  \begin{minipage}[c]{.46\linewidth}
\include{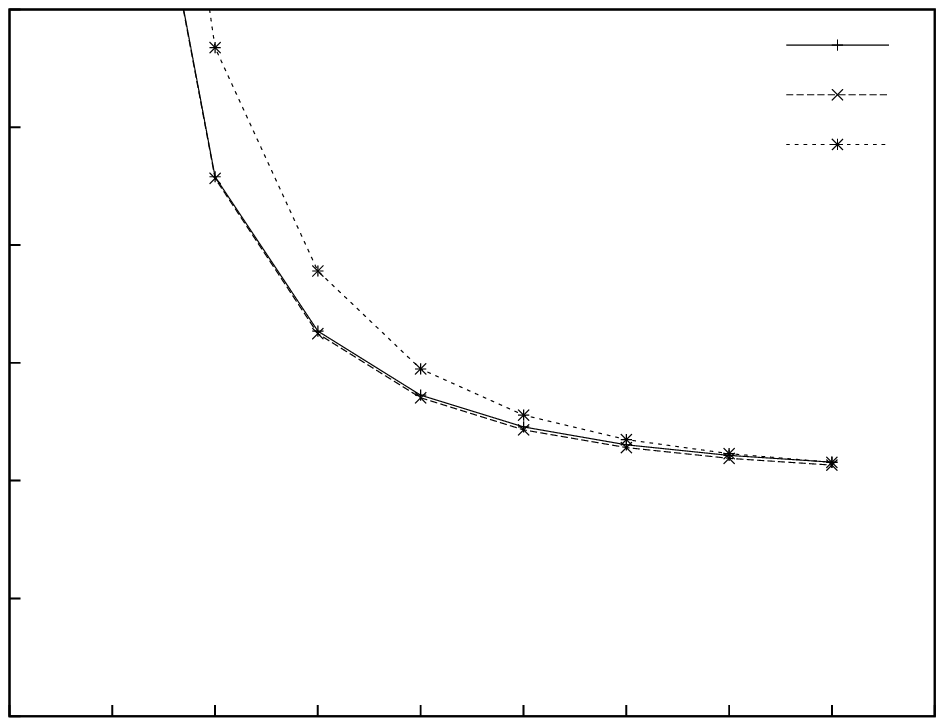}
   \end{minipage} \hfill
   \begin{minipage}[c]{.46\linewidth}
\include{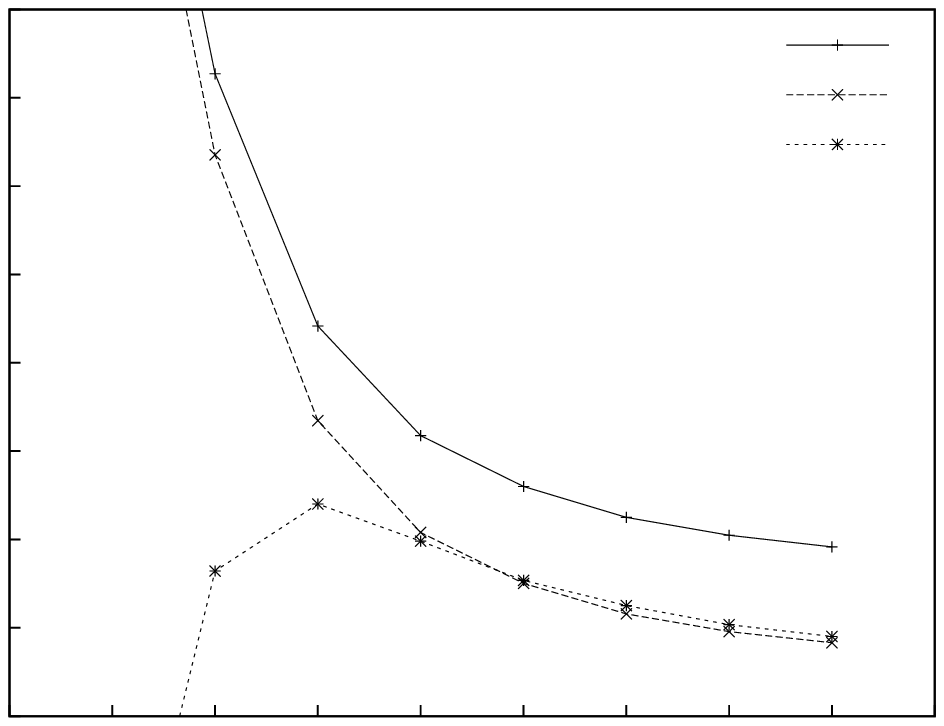}
   \end{minipage}
\include{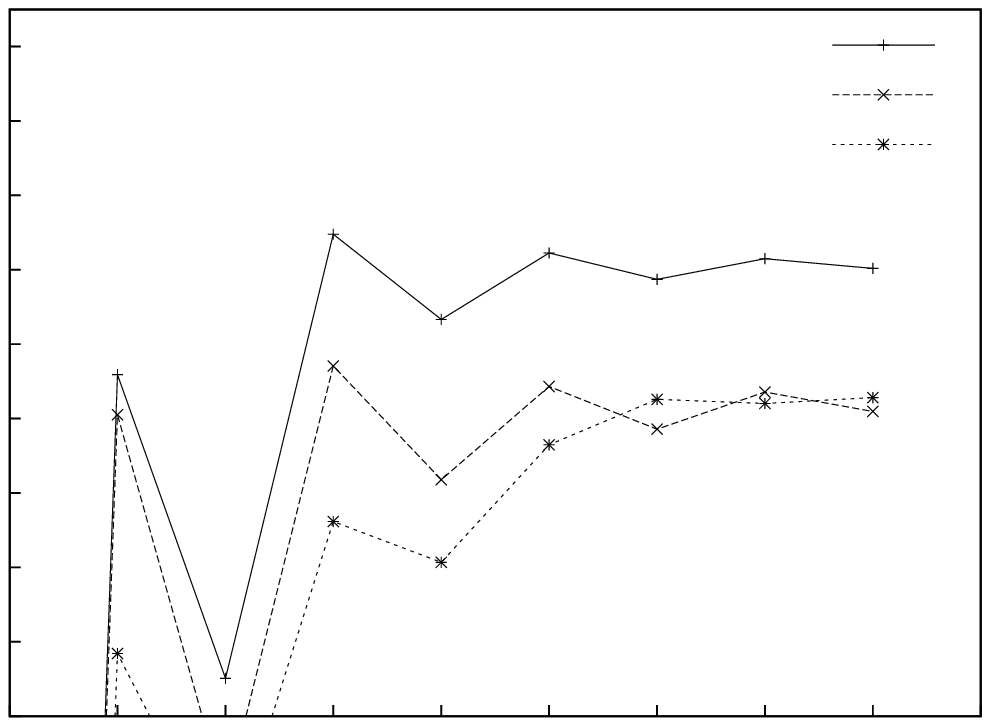}
\end{center}
\caption{Convergence of the absolute total energy, the specific mass shift and the hyperfine contact parameters for the ground state of neutral lithium. The agreement found between the SD-MCHF and SD-PCFI curves for the total energy is not observed for the two other properties. We recover the consistency with the traditional method by deconstraining the wave function (SD-DPCFI).}
\label{fig:cont_comp}
\end{figure}

We compare the expectation values of other operators than the Hamiltonian, i.e. the specific mass shift and the hyperfine interaction parameters~\cite{Godetal:01a}, evaluated by the two SD-(mono)-PCFI and SD-(mono)-MCHF methods in \fref{fig:cont_comp}. The two curves illustrate  the impact of the ``constraint effect'' on three different properties: the  total energy, the specific mass shift ($S_{\textrm{sms}}$) and the contact term ($a_{\textrm{cont}}$).
As mentioned in the end of \sref{PCFI}, the origin of this effect is the hindrance to free variation in the expansion coefficients and in the orbitals. Even if the pre-optimized PCF orbital sets are fixed in the PCFI diagonalization step, the associated orbital constraint effect is expected to be small in comparison to the constraint on the mixing coefficients.
The lack of variation in the orbitals is indeed somewhat compensated by the use of separated PCF orbital sets that makes the number of radial functions for a given active set, larger in the PCFI approach than in the MCHF method.  Moreover the number of correlation layers used for a given PCF is probably large enough to reach saturation. The strongest limitation is likely to come from the fact that 
for each PCF, the mixing coefficients  appearing in~\eref{eq:Lambda_frozen} are kept frozen in the interaction step that leads to the final many-electron wave function expression~\eref{eq;wfn}. By freezing these mixing coefficients, we inhibit the expression of any indirect effects. The relative weights of the CSFs within each PCF are indeed already fixed by the frozen coefficients $\{ c_j^\Lambda \}$ obtained from each independent MR-PCF MCHF optimization of \eref{eq:MR-PCF}, each one targeting a specific correlation component,  and no possibility is offered to these coefficients to capture the higher-order PCF-coupling effects. \\

Engels~\cite{Eng:93a} studied the influence of various excitation classes on ab initio calculated isotropic hyperfine constants and showed how each class gives a direct contribution, but that there are also important secondary effects on the contributions from the other classes due to relative changes of mixing coefficients.
The importance of 
these indirect effects on the CSF weights that affect computed properties quite substantially is confirmed in the present work. Back to the beryllium example, \Tref{tab:Be1P-1S_trans} that collects the radiative data for the resonance E1 transition of beryllium illustrates that the constraint effect can also be significant for properties involving more than one state. Comparing the $gf$ values from the SD-MR-MCHF and SD-MR-PCFI calculations we see that the latter converges much more rapidly, but that there is an off-set of a little less than 1~\%. Also, the agreement between the length and velocity forms is not as good for the PCFI method as for the ordinary MCHF. 


\begin{table}[!ht]
\caption{\label{tab:Be1P-1S_trans}Line and oscillator strength of the $^1P^o-\, ^1S$ resonant line for the beryllium atom.}
\begin{indented}
\item[]\begin{tabular}{ l | l | l l | l l }
\hline
\multicolumn{6}{c}{SD-MR-MCHF} \\ \hline 
$n_{max}$ & $\Delta \textrm{E}$ (cm$^{-1}$)  & $S_{l}$ & $S_{v}$ & $gf_{l}$ & $gf_{v}$ \\
\hline
4  & $43513.26$ & $10.6832$ & $10.9355$ & $1.41204$ & $1.44539$  \\
5  & $43069.99$ & $10.6269$ & $10.7335$ & $1.39029$ & $1.40424$  \\
6  & $42843.85$ & $10.6060$ & $10.6049$ & $1.38027$ & $1.38013$  \\
7  & $42686.81$ & $10.6110$ & $10.6255$ & $1.37587$ & $1.37774$  \\
8  & $42635.54$ & $10.6205$ & $10.6301$ & $1.37544$ & $1.37668$  \\
9  & $42606.57$ & $10.6249$ & $10.6366$ & $1.37508$ & $1.37659$  \\
10 & $42593.62$ & $10.6273$ & $10.6375$ & $1.37497$ & $1.37628$  \\
\hline
\multicolumn{6}{c}{SD-MR-PCFI $\{\underline{1s},\underline{2s},\underline{2p},\underline{3s},\underline{3p},\underline{3d}\}$} \\ \hline 
$n_{max}$ & $\Delta \textrm{E}$ (cm$^{-1}$)  & $S_{l}$ & $S_{v}$ & $gf_{l}$ & $gf_{v}$ \\
\hline
4  & $42731.10$ & $10.6848$ & $10.6446$ & $1.38686$ & $1.38165$  \\
5  & $42649.36$ & $10.7012$ & $10.7249$ & $1.38634$ & $1.38940$  \\
6  & $42618.07$ & $10.7047$ & $10.7347$ & $1.38578$ & $1.38966$  \\
7  & $42603.40$ & $10.7083$ & $10.7391$ & $1.38576$ & $1.38975$  \\
8  & $42596.17$ & $10.7096$ & $10.7412$ & $1.38570$ & $1.38979$  \\
9  & $42591.03$ & $10.7104$ & $10.7433$ & $1.38563$ & $1.38988$  \\
10 & $42588.21$ & $10.7110$ & $10.7452$ & $1.38561$ & $1.39004$  \\
\hline
\multicolumn{6}{c}{SD-MR-DPCFI $\{\underline{1s},\underline{2s},\underline{2p},\underline{3s},\underline{3p},\underline{3d}\}$} \\ \hline 
$n_{max}$ & $\Delta \textrm{E}$ (cm$^{-1}$)  & $S_{l}$ & $S_{v}$ & $gf_{l}$ & $gf_{v}$ \\
\hline
4  & $42725.56$ & $10.6443$ & $10.5604$ & $1.38143$ & $1.37055$  \\
5  & $42640.89$ & $10.6355$ & $10.6149$ & $1.37755$ & $1.37489$  \\
6  & $42608.81$ & $10.6324$ & $10.6235$ & $1.37612$ & $1.37497$  \\
7  & $42593.58$ & $10.6358$ & $10.6263$ & $1.37606$ & $1.37483$  \\
8  & $42586.00$ & $10.6361$ & $10.6287$ & $1.37586$ & $1.37491$  \\
9  & $42580.53$ & $10.6363$ & $10.6301$ & $1.37570$ & $1.37490$  \\
10 & $42577.57$ & $10.6369$ & $10.6322$ & $1.37569$ & $1.37508$  \\
\hline
CAS ($n_{max} l_{max} =9 \ell) $ & $42588.71$ & $10.6234$ & $10.6333$ & $1.37431$ & $1.37559$ \\
  & & & & & \\
Chung and Zhu~\cite{ChuZhu:93a} & $42559.20$ & $$ & $$ & $1.374$ & $$ \\
Fleming {\it et al.}~\cite{Fleetal:96c} & $42593.44$ & $$ & $$ & $1.3745$ & $1.3759$\\
Komasa and Rychlewski~\cite{KomRyc:2001a} & $42560.52$ & $$ & $$ & $$ & $$ \\
G\l owacki and Migda\l ek~\cite{GloMig:2006a} & $42670.69$ & $$ & $$ & $1.340$ & $$\\
  & & & & & \\
Kramida and Martin~\cite{KraMar:97a} & $42565.35$ & & & & \\
Irving {\it et al.}~\cite{Irvetal:99a} & & &  &  $1.40 \pm 0.04$ & \\
Schnabel and Kock~\cite{SchKoc:2000a} & $$ & & &$1.34\pm0.03$ &  \\
\end{tabular}
\end{indented}
\end{table}

\clearpage

\section{Deconstraining Partitioned Correlation Functions}

In the PCFI method, the expansion coefficients for the CSFs in the PCF are constrained (locked) so that there is no possibility of relative changes due to the interaction with other PCFs. To recover this variational freedom, the PCFs can be deconstrained by transferring $h_j$ CSFs from the $j$-th PCF  to the basis and at the same time setting their weights to zero, i.e. extending the PCFI space to
\begin{equation} 
\hspace*{-1cm} 
\{ \Phi^{\textsc{mr}}_1 , \ldots,  \Phi^{\textsc{mr}}_m \} \bigcup_{j=1}^p 
\{ {\overline{\Lambda}}_j \}
\longrightarrow
\{ \Phi^{\textsc{mr}}_1 , \ldots,  \Phi^{\textsc{mr}}_m \} \bigcup_{j=1}^p 
\{ \Phi^{j}_1 , \ldots,  \Phi^{j}_{h_j},
\overline{\Lambda}_j^{d} \}.
\label{eq:transfer}
\end{equation}
The superscript $d$ for the PCF $\overline{\Lambda}_j^{d}$ indicates a renormalized {\it de}-constrained Partitioned Correlation Function whose weights of the transferred CSFs have been set to zero.  The many-electron wave function expansion becomes
\begin{equation}
\label{eq:DPCF}
\Psi = \sum_{i=1}^m a_i  \Phi^{\textsc{mr}}_i  + \sum_{j=1}^p \left\{ \sum_{k=1}^{h_j} c^{j}_k \Phi^{j}_{k}  + {c}_j {\overline{\Lambda}}_j^{d}  \right\},
\end{equation} 
where the expansion coefficients are obtained from a higher dimension ($M = m + \sum_{j=1}^p (h_j + 1)$) {\it a priori} generalized eigenvalue problem. The size of each block in~\eref{eq:big_mat} involving at least one PCF  is growing accordingly to the number of deconstrained CSFs. 
In the limit of the completely deconstrained case (ie. $h_j = dim ( {\overline{\Lambda}}_j $) $ \; \forall j$ ), we regain full variational freedom in the coefficients, with the advantage that each CSF brings its tailored orbital basis.
 $M$ is then at his maximum value, ie. the total number of CSFs, and the wave function~\eref{eq:DPCF} will be referred as being ``deconstrained''. It is strictly equivalent to a configuration interaction problem in the CSF space built on various mutually non-orthonormal orbital sets. \\

For solving the eigenvalue problem \eref{GEP} and for building efficiently the interaction matrices associated with the selected operators in the basis of CSFs and deconstrained PCFs spanning the wave function~\eref{eq:DPCF}, we modify the original way of presenting the biorthonormal transformation~\cite{Olsetal:95a} to evaluate $O_{ij} = \langle {\Lambda}_i  | O | {\Lambda}_j \rangle$, where $O$ is the Hamilton or unit operator. Following the original formalism,  we perform a biorthonormal transformation 
\[
\langle \phi^i_k | \phi^j_m \rangle = S_{km}^{ij} \rightarrow \langle \tilde{\phi}^i_k | \tilde{\phi}^j_m \rangle = \delta_{n_k,n_m}\delta_{l_k,l_m} 
\]
to express the original left and right hand side PCFs in the new CSF bases $\{  \tilde{\Phi}^{i}_k  \}$ and $ \{ \tilde{\Phi}^{j}_l \}$
\begin{equation}
| \Lambda_i \rangle =  \sum_{k=1}^{n_i} d^i_k | \Phi^{i}_k \rangle = \sum_{k=1}^{n_i} \tilde{d}^i_k | \tilde{\Phi}^{i}_k \rangle
\end{equation}
\begin{equation}
| \Lambda_j \rangle =  \sum_{l=1}^{n_j} d^j_l | \Phi^{j}_l \rangle = \sum_{l=1}^{n_j} \tilde{d}^j_l | \tilde{\Phi}^{j}_l \rangle \;,
\end{equation}
where the counter-transformed eigenvectors $\{  \tilde{d}^i_k \}$ and $\{  \tilde{d}^j_l \}$ ensure the invariance of the total wave functions. Given the matrix representation  $\tilde{{\bf O}}$ of an operator $O$
\begin{equation}
\tilde O_{kl} = \langle \tilde{\Phi}^{i}_k | O | \tilde{\Phi}^{j}_l \rangle\;,
\label{eq:mat_operator}
\end{equation} 
the matrix element between these PCFs is written as
\begin{equation}
\langle {\Lambda}_i  | O | {\Lambda}_j \rangle = \sum_{k,l}  (\tilde{d}^i_k)^*  \tilde{d}^j_l \langle \tilde{\Phi}^{i}_k | O |  \tilde{\Phi}^{j}_l \rangle = (\tilde{{\bf d}}^{i})^t \tilde{{\bf O}}\tilde{{\bf d}}^{j}
\label{eq:block}
\end{equation}
where $\tilde{{\bf d}}$ is the column vector of counter-transformed mixing coefficients. Note that each CSF expansion should be {\it closed under de-excitation} (CUD) for allowing the biothonormal transformation~\cite{Olsetal:95a,Veretal:2010a}. By strictly following this methodology, we may think that we should apply one biorthonormal transformation for each matrix element associated with any off-diagonal sub-matrix block involving two non-orthogonal orbital sets. 
However, as the counter-transformation process is fixed by the overlap between the original spin-orbital bases, i.e. $\langle \phi^i_k | \phi^j_m \rangle = S_{km}^{ij}$, we show in the Appendix~1 that one can evaluate the whole sub-matrix block by performing a {\it single} biorthonormal transformation treating simultaneouly the counter-transformation of all the elements constituting the block-basis. \\

Using this strategy, relaxing the PCFI constraint to any degree becomes possible. The price to pay is the increase of the size of the Partition Correlation Function Interaction problem \eref{GEP}. In the limit of the completely deconstrained case (ie. $h_j = dim ( {\overline{\Lambda}}_j )\; \;  \forall j$) , the PCFI approach is strictly equivalent to a configuration interaction problem in the original complete CSF space. We will use the label DPCFI for this ``deconstrained'' approach. Since each CSF could be built, if worthwhile, on its own orbital basis, without any radial orthogonality constraints with the other ones, this DPCFI approach is equivalent to a general configuration interaction problem in non-orthogonal orbitals. \\

\noindent The (D)PCFI procedure can be summarized as follows: 
\begin{enumerate}
\item Perform a HF/MCHF calculation for the mono-/multi-reference wave function~\eref{eq:MR},
\item Freeze the orbitals belonging to this MR space and perform $p$ separate MR-PCF MCHF calculations~\eref{eq:MR-PCF} for the different Partitioned Correlation Functions,
\item (Optional) Deconstrain each PCF by transferring the desired CSFs from the CF to the MR basis (see \eref{eq:transfer}),
\item Build the Hamiltonian and other relevant operators interaction matrices~\eref{eq:big_mat} by performing the biorthonormal transformations, if necessary, using the weight matrix formalism (see Appendix~2), 
\item Solve the eigenvalue problem~\eref{GEP} for getting the many-electron wave function~\eref{eq;wfn},
\item Compute the desired property with the PCFI eigenvector(s).
\end{enumerate}
As it is shown in Appendix~1, step (ii) allows to replace the overlap matrix equal of \eref{GEP} by the unit matrix.

\clearpage
 
\section{Applications of the DPCFI method}
\label{sec:Appl-DPCFI}

\subsection{Solving the constraint problem}

This ``Deconstrained Partitioned Correlation Function Interaction'' (DPCFI) strategy  has been applied to  both lithium and beryllium  by solving the eigenvalue problem in the original CSF basis instead of the PCF one. Each CSF comes with the orbital basis associated with the PCF from which it comes from.
 In both cases, as illustrated by \fref{fig:Be_1P} and \fref{fig:cont_comp} 
for beryllium and lithium respectively, the DPCFI values converge to the MCHF results, recovering all the indirect effects and keeping the tremendous advantage of a much faster convergence. This corroborates the fact that the problematic PCFI-MCHF discrepancy observed for the  $S_{\mathrm{sms}}$ and contact hyperfine parameters is due to the constraint effect. 
  On the other hand, total energies are only slightly improved in the (PCFI $\rightarrow $ DPCFI) transition. 
A close analysis of the results displayed in \Fref{fig:cont_comp} shows that for $n=10$,
the relative differences (in absolute value) between the PCFI and DPCFI results are respectively $1.65~10^{-4}~\%$, $0.18~\%$ and $1.33~\%$ for the  total energy, the specific mass shift and the contact term parameters. These numbers demonstrate that, compared with the total energy, the specific mass shift and the contact term are much more sensitive to the constraint effect.  \\

\subsection{The beryllium resonance line}
\label{subsec:Be_DPCFI}

For the beryllium resonance transition (see~\tref{tab:Be1P-1S_trans}),  passing from SD-MR-PCFI to SD-MR-DPCFI, the line strength is modified by about $0.7~\%$ in the length formalism and $1.1~\%$ in the velocity formalism, bringing the two forms in much closer agreement. The SD-MR-DPCFI is now in excellent agreement with the most extensive MCHF calculations available ($n = 9$ CAS).
 As the absolute energy of each level, the transition energy is much less sensitive to the constraint effect since it undergoes a modification of less than $0.02~\%$. Note that both the SD-MR-PCFI and the SD-MR-DPCFI methods provide a transition energy at the $n=6$ level which is comparable to the SD-MR-MCHF $n=10$ result. As already pointed out in our first paper~\cite{Veretal:2010a}, the richness in terms of radial functions of the many-electron wave functions adopting different PCF definitely leads to a higher rate of convergence for all properties. \\

It is probably worthwhile to review briefly the status of the available oscillator strength values for this Be~I resonance line reported in~\tref{tab:Be1P-1S_trans}. Experimentally, the situation has evolved since the publication in 1996 of the theoretical results of Fleming~{\it et al.}~\cite{Fleetal:96c} of $gf_l = 1.3745$ in the length form and $gf_v = 1.3759$ in the velocity form. Irving~{\it et al.}~\cite{Irvetal:99a} revised the old beam-foil measurement of Martinson~{\it et al.}~\cite{Maretal:74a}, by including the cascade corrections through the ANDC analysis that increased the oscillator strength from $gf = 1.34 \pm 0.05 \rightarrow 1.40 \pm 0.04$, in nice agreement with the theoretical predictions. An independent measurement has been realized by Schnabel and Kock~\cite{SchKoc:2000a} using the cascade-free laser induced fluorescence method, yielding the original beam-foil result but with a smaller uncertainty, namely $1.34 \pm 0.03$. Amazingly, this value escaped to the attention of the authors of a rather complete compilation~\cite{TraCur:06a}. 
\Tref{tab:Be1P-1S_trans} shows that the present SD-MR-DPCFI results confirm the previous theoretical value~\cite{Fleetal:96c}, with a difference between the two gauges that is much smaller than the accuracy indicators reported by Fuhr and Wiese~\cite{FuhWie:2010a}. The small overlap between the two most recent experimental values~\cite{Irvetal:99a,SchKoc:2000a} call for further investigations on the experimental side.\\

\subsection{The CAS-DPCFI approach in lithium}
In this section we present results obtained for selected spectroscopic properties involving the two lowest states of neutral lithium, ie. $1s^2 2s \; ^2S$ and $1s^2 2p \; ^2P^o$. The Li~I ground state (D)PCFI calculations that were presented in \sref{constraint} and \fref{fig:cont_comp} to investigate the constraint effect are  limited in the sense that triple excitations are systematically omitted in a SD-mono-reference calculation.
In order to  obtain more accurate  many-electron wave functions, we adopted the complete active space (CAS) list of CSFs combined with the DPCFI approach that remains manageable for a three-electron system.  
For illustrating the great flexibility of the DPCFI approach, we investigate two different strategies - i) treating  core excitations globally and ii) separating the single and double core-excitations and dedicating a PCF to core-polarization.
\subsubsection{A global core description}
Like in our first model, we use the Hartree-Fock (HF) solution as the zeroth-order wave function for the two lowest states of this three electron system. In this approach, we split the CF space made of single, double and triple excitations (SDT) in three   different subspaces ($p=3$) defining the following three PCFs: 
\begin{itemize}
\item one for taking care of the inner-shell correlation between the two $1s$ electrons
\begin{eqnarray}
| \Lambda_{1s-1s1s} \rangle &=& | 1s^22l~^2L^{\pi} \rangle + \sum_{n'l'} | 1s2l\;n'l'~^2L^{\pi} \rangle \nonumber\\
& & + \sum_{n'l',n''l''} | 2l\;n'l' n''l''~^2L^{\pi} \rangle \;, \label{eq:Lamnda1s-1s1s}
\end{eqnarray}

\item a second one associated to the inter-shell correlation between the $1s$ and the $2l$ ($l=s\textrm{ or } p$) 
\begin{eqnarray}
| \Lambda_{2l-1s2l} \rangle &=& | 1s^22l~^2L^{\pi} \rangle + \sum_{n'l'} | 1s^2\;n'l'~^2L^{\pi} \rangle \nonumber\\
& & + \sum_{n'l',n''l''} | 1s\;n'l' n''l''~^2L^{\pi} \rangle \;, \label{eq:Lamnda2l-1s2l}
\end{eqnarray}

\item a third and last one including the pure triple excitations
\begin{eqnarray}
| \Lambda_{1s1s2l} \rangle &=& | 1s^22l~^2L^{\pi} \rangle + \sum_{n'l',n''l'',n'''l'''} | n'l' n''l''n'''l'''~^2L^{\pi} \rangle \;. \label{eq:Lamnda1s1s2l}
\end{eqnarray}
\end{itemize}
The above notation for the three PCFs applies to both $1s^2 2s \; ^2S$  or $1s^2 2p \; ^2P^o$ states, 
with $L=l$.  \\

\begin{figure}[!ht]
  \begin{minipage}[c]{.46\linewidth}
\include{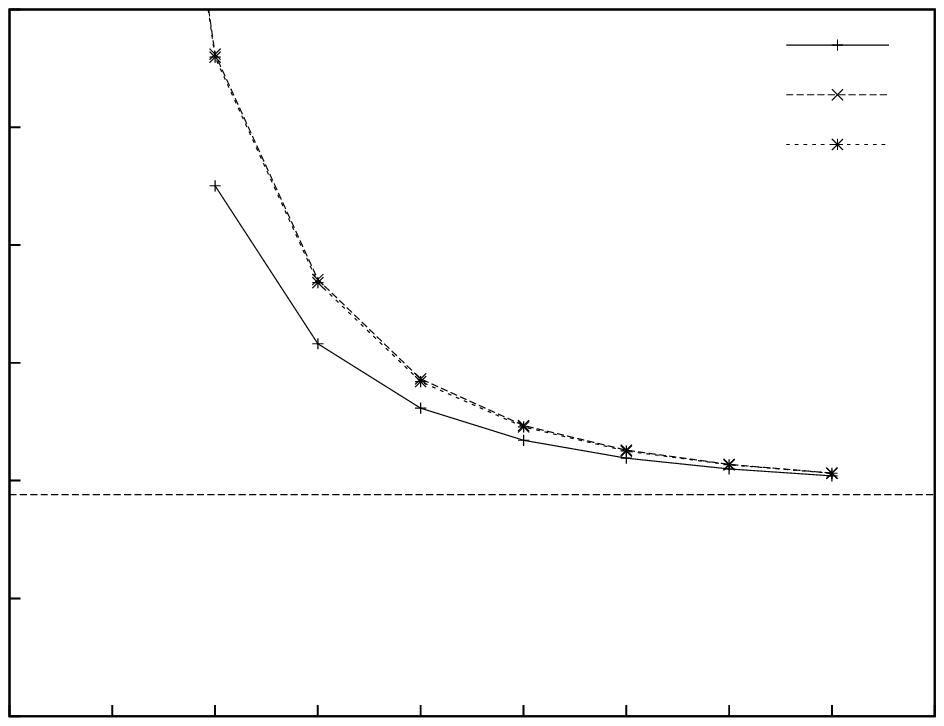}
   \end{minipage} \hfill
   \begin{minipage}[c]{.46\linewidth}
\include{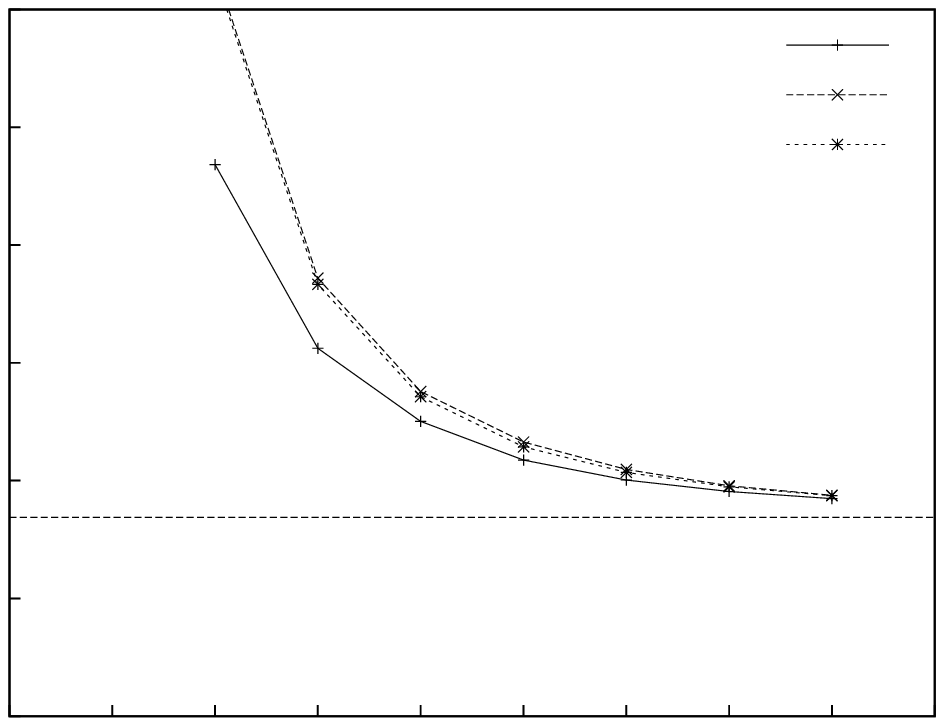}
   \end{minipage}
\caption{Convergence of the total energy with respect to the largest principal quantum number of the active sets for the ground state $1s^2 2s \; ^2S$ of lithium on the left and for the first excited state $1s^2 2p \; ^2P^o$ on the right. The reference values (dotted lines) correspond to the absolute energy values obtained by Yan \emph{et al.} \cite{YanDra:95a}.}
\label{fig:1s1s-1s2l-Triple_E}
\end{figure}

\begin{figure}[!ht]
  \begin{minipage}[c]{.46\linewidth}
\include{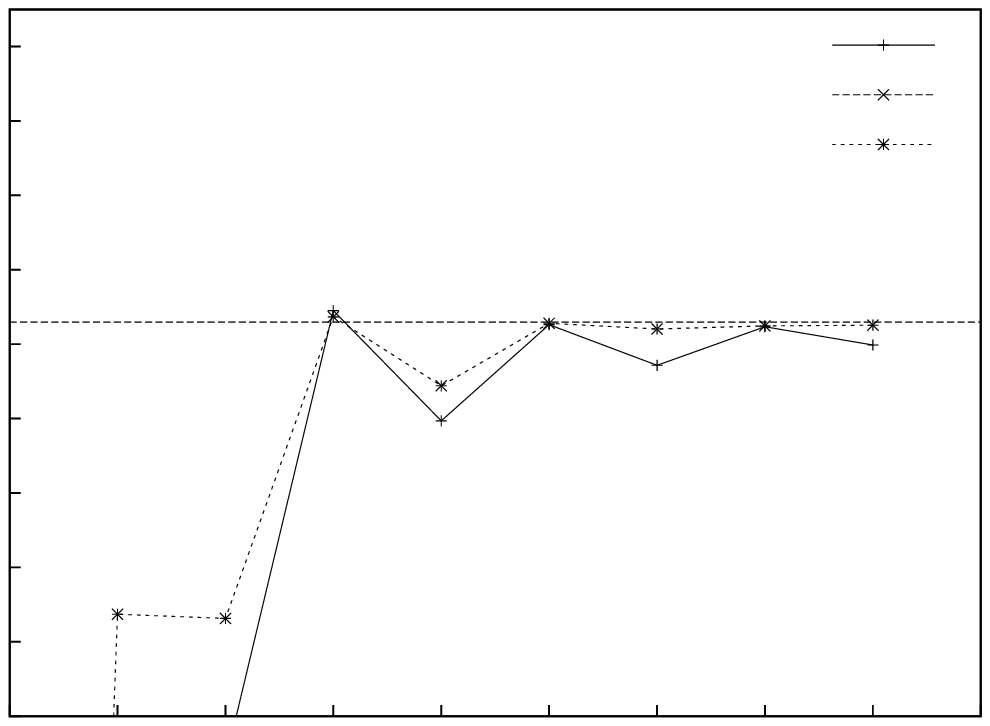}
   \end{minipage} \hfill
   \begin{minipage}[c]{.46\linewidth}
\include{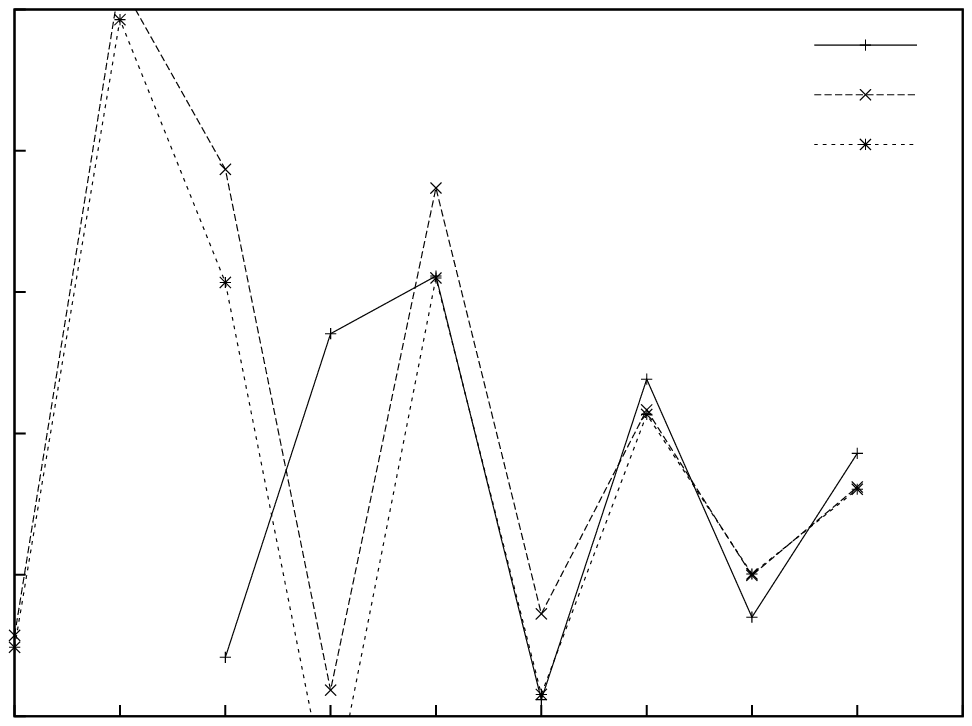}
   \end{minipage}
\caption{Convergence of the hyperfine contact parameter of the ground state (on the left) and of the electric quadrupole hyperfine parameter   of the first excited state (on the right) with respect to the largest principal quantum number of the active sets. The reference value for the contact $a_c$ parameter (dotted line) corresponds to the value obtained by Yan \emph{et al.} \cite{Yanetal:96a}.}
\label{fig:1s1s-1s2l-Triple_hfs}
\end{figure}

 For the first two PCFs \eref{eq:Lamnda1s-1s1s} and \eref{eq:Lamnda2l-1s2l},   we optimize all the correlation orbitals, freezing the $1s$ and $2l$ ($= 2s$ {\it or} $2p$) to the HF solution of the mono-reference. This strategy is  inadequate for $\Lambda_{1s1s2l}$ since it only contains triple excitations that do not interact with the reference CSF.  The optimization of the corresponding orbital set becomes then more tricky. In the previous paper~\cite{Veretal:2010a}, we chose to use a SD-multireference to include triple excitations. 
 In the present work, we dedicate a specific PCF to these.  We first define an ``extended'' SD expansion for a reference set built on the $(n=2,3)$ shells  and optimize it by allowing variations in the correlation orbitals only. This expansion opens an indirect interaction between the triple excitations and the reference CSF.   For capturing these higher-order  effects, we optimize the $n > 3$ orbitals during the  MR-PCF procedure. The $n=4$ layer is therefore the first one that effectively represents three-electron excitations. 
\Fref{fig:1s1s-1s2l-Triple_E}  illustrates for the two states that the DPCFI convergence is faster than the traditional CAS-MCHF approach based on the same CSF expansions. For a given orbital active set, the corresponding total energy value is indeed systematically below the CAS-MCHF result. Since the angular content of the wave function (maximum $l$-value for in the one-electron basis) is identical for both methods, we conclude that the DPCFI method captures more efficiently electronic correlation for a given atomic system. 
A similar improvement is a priori expected for any other spectroscopic property. \Fref{fig:1s1s-1s2l-Triple_hfs}  presents the convergence pattern of the contact term for the ground state and the electric quadrupole parameter of the first excited state of neutral lithium. As it clearly appears, the hyperfine parameters are not converging as smoothly as the total energy. It is well known that the relevant expectation values are extremely sensitive to single excitations~\cite{Eng:93a} and it is worthwhile to attempt another approach for treating this excitation family independently.

\subsubsection{A Partitioned Correlation Function dedicated to core-polarization} 
For describing more accurately the hyperfine interaction, we split the $\Lambda_{1s- 1s1s}$ PCF in two subspaces
\begin{equation}
\Lambda_{1s- 1s1s} \rightarrow \Lambda_{1s} + \Lambda_{1s1s} \; . \label{eq:split}
\end{equation}
 The $\Lambda_{1s}$  PCF function focusing on the single excitations is dedicated to capture core-polarization effects. The many-electron wave function is then written as the reference function corrected by four ($p=4$) different PCFs with their own orbital set: the two first, associated to the core-valence~\eref{eq:Lamnda2l-1s2l} and to the triple excitations~\eref{eq:Lamnda1s1s2l}, as described in the previous subsection, completed by
\begin{itemize}
\item a third one that takes care of the single excitations of the $1s$ shell
\begin{eqnarray}
| \Lambda_{1s} \rangle &=& | 1s^22l~^2L^{\pi} \rangle + \sum_{n'l'} | 1s2l\;n'l'~^2L^{\pi} \rangle \;, \label{eq:Lamnda1s}
\end{eqnarray}

\item a last one associated to the double excitations of the $1s$ shell 
\begin{eqnarray}
| \Lambda_{1s1s} \rangle &=& | 1s^22l~^2L^{\pi} \rangle + \sum_{n'l',n''l''} | 2l\;n'l' n''l''~^2L^{\pi} \rangle  \label{eq:Lamnda1s1s}\;.
\end{eqnarray}
\end{itemize}

\begin{figure}[!ht]
  \begin{minipage}[c]{.46\linewidth}
\include{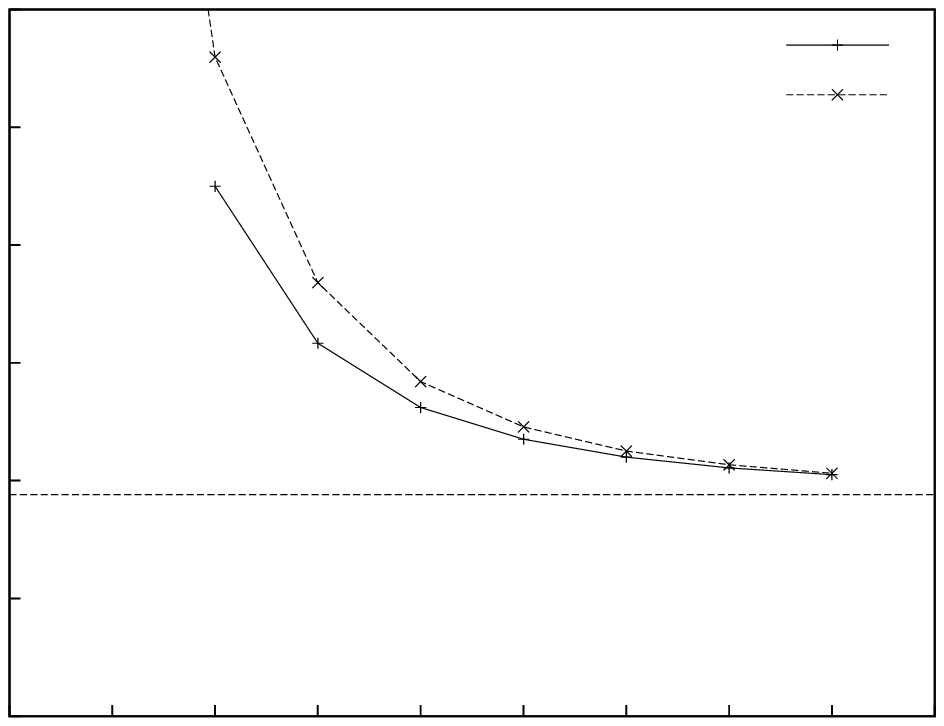}
   \end{minipage} \hfill
   \begin{minipage}[c]{.46\linewidth}
\include{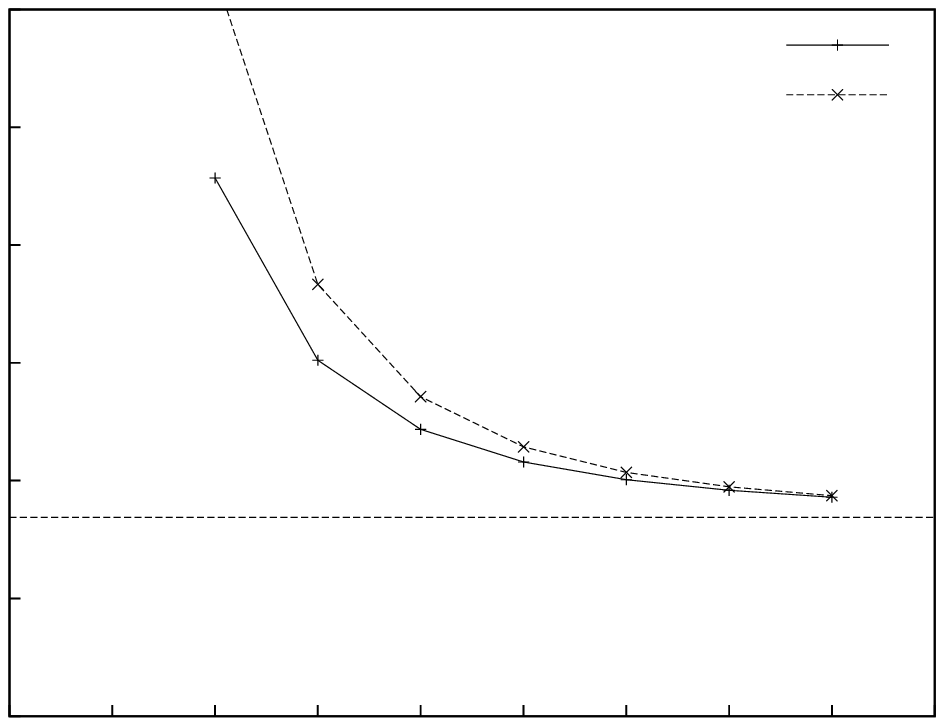}
   \end{minipage}
\caption{Convergence of the absolute energy with respect to the largest principal quantum number of the active sets for  $1s^2 2s \; ^2S$ (left) and $1s^2 2p \; ^2P^o$ (right) in neutral lithium. The reference values (dotted line)  correspond to the results obtained by Yan \emph{et al.} \cite{YanDra:95a}.}
\label{fig:1s-1s1s-1s2l-Triple_E}
\end{figure}

\begin{figure}[!ht]
\begin{center}
\include{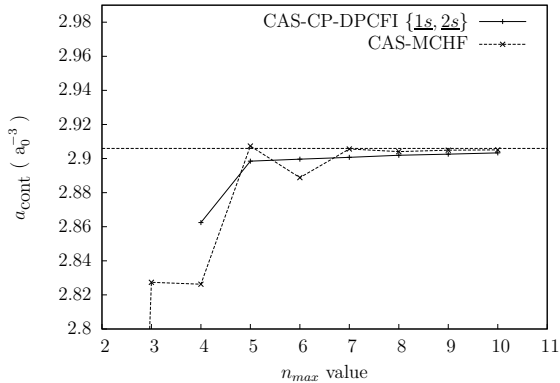}
\caption{Convergence of the hyperfine contact parameter  with respect to the largest principal quantum number of the active sets for the ground state of lithium. The reference value (dotted line) correspond to the results obtained by Yan \emph{et al.} \cite{Yanetal:96a}.}
\label{fig:1s-1s1s-1s2s-Triple_hfs}
\end{center}
\end{figure}

\begin{figure}[!ht]
  \begin{minipage}[c]{.46\linewidth}
\include{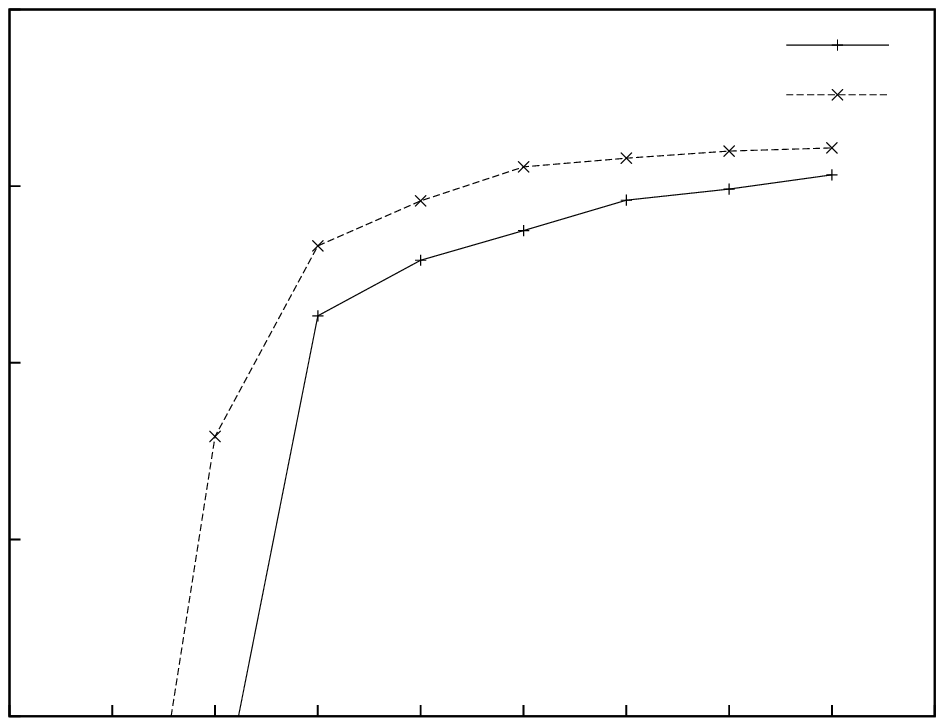}
   \end{minipage} \hfill
   \begin{minipage}[c]{.46\linewidth}
\include{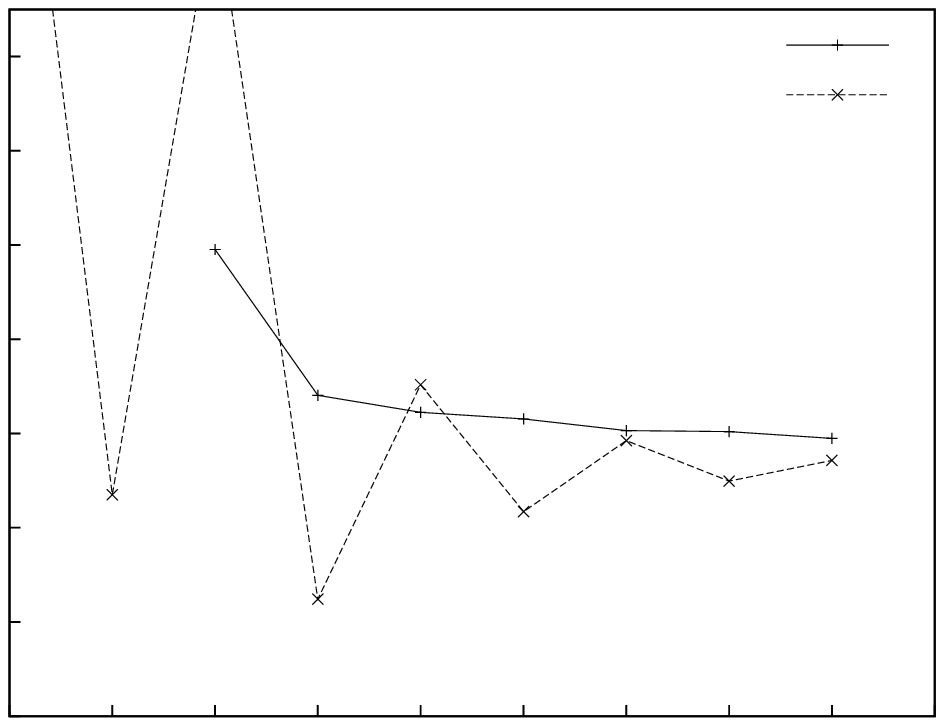}
   \end{minipage}
     \begin{minipage}[c]{.46\linewidth}
\include{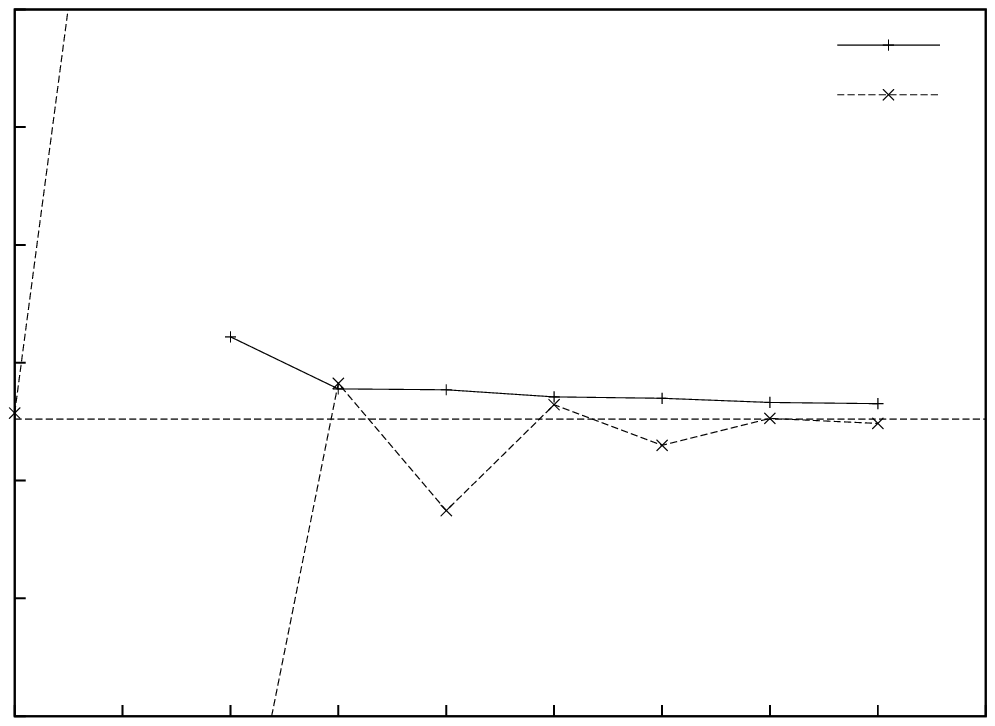}
   \end{minipage} \hfill
   \begin{minipage}[c]{.46\linewidth}
\include{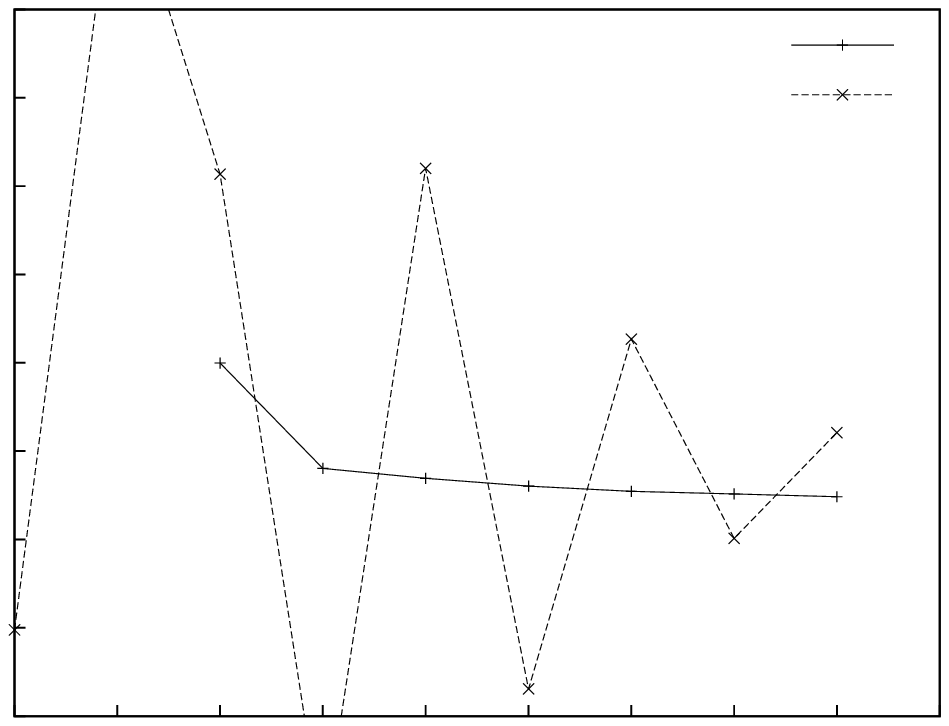}
   \end{minipage}
\caption{Convergence of the hyperfine parameters with respect to the largest principal quantum number of the active sets for the first excited state, $1s^2 2p \; 2P^o$ of lithium. The reference value (dotted line) corresponds to the result obtained by Yan \emph{et al.} \cite{Yanetal:96a}.}
\label{fig:1s-1s1s-1s2p-Triple_hfs}
\end{figure}

Excitations considered in \eref{eq:Lamnda1s} describe spin-polarization, for both the $^2S$ and $^2P^o$ states, since single excitations can break the singlet spin coupling between the two core electrons. For the first excited state, these excitations can also break the angular coupling associated with orbital-polarization. It is well known that the hyperfine parameters are sensitive to these excitations~\cite{LinMor:82a} and some improvement is expected in their evaluation thanks to the splitting \eref{eq:split}. The results are presented in \fref{fig:1s-1s1s-1s2l-Triple_E}, \fref{fig:1s-1s1s-1s2s-Triple_hfs} and \fref{fig:1s-1s1s-1s2p-Triple_hfs}. 
By comparing \fref{fig:1s1s-1s2l-Triple_E} and \fref{fig:1s-1s1s-1s2l-Triple_E}, it is obvious that the decomposition~\eref{eq:split}  does not affect the total energy value. The DPCFI method still captures correlation more efficiently than the traditional MCHF calculations.  
The interesting improvement appears for the different hyperfine parameters. \Fref{fig:1s-1s1s-1s2s-Triple_hfs} and \fref{fig:1s-1s1s-1s2p-Triple_hfs} illustrate their progressive convergence, respectively for the ground and the first excited states. The use of the orbital set tailored for capturing the spin- and orbital core-polarization (CP) enhanced beautifully the convergence pattern of all the hyperfine parameters. The resulting trends are much smoother than those of the global core approach and the ordinary MCHF (see \fref{fig:1s1s-1s2l-Triple_hfs}). All oscillations disappeared and we reach reasonably well-converged values around $n=5$.
Lithium is a small atom and it is possible to enlarge an ordinary orbital basis to get converged  values for all quantities. For larger atoms with more complicated shell structures it is, to set 
things into perspective, often not possible to extend the radial orbital basis very much due to a rapidly growing number of CSFs and here the fast convergence of the CP-DPCFI method, together with the fact that orbital sets for different shells can be optimized independently of each other, represents a {\em major} improvement in the general methodology.


The values of the corresponding total energies, $S_{\textrm{sms}}$ and hyperfine parameters are reported in tables~\ref{tab:2S_SP+T} and \ref{tab:2P_SP+T}.
As in the figures, we compare the DPCFI values  with  Yan \emph{et al.}~\cite{YanDra:95a,Yanetal:96a} results using Hylleraas-type variational method. The remaining difference between both studies may be attributed to the slow angular convergence rate ($E_l-E_{l-1} = \mathcal{O}(l+1/2)^{-4}$) of the $(1/r_{12})$ angular development~\cite{Hil:85a}. 
\begin{table}[!ht] 
\caption{\label{tab:2S_SP+T}Energies, $S_{\textrm{sms}}$ and $a_{\textrm{cont}}$ for an increasing maximum principal quantum number for the ground state of Li.}
\begin{indented}
\item[]\begin{tabular}{c | l l l}
\hline
\multicolumn{4}{c}{CAS-CP-DPCFI $\{\underline{1s},\underline{2s}\}$ } \\
\hline
$n_{max}$ & Energy (a.u.) & $S_{\textrm{sms}}$ & $a_{\textrm{cont}}$\\
\hline
HF  	&$-7.432726927$  &$0.000000000$ &$2.0932317$ \\
4 & $-7.476750919$ & $0.304445112$  & $2.8624693$  \\
5 & $-7.477417364$ & $0.303171470$  & $2.8984561$  \\
6 & $-7.477689394$ & $0.302588400$  & $2.8996669$  \\
7 & $-7.477824739$ & $0.302322528$  & $2.9007289$  \\
8 & $-7.477900177$ & $0.302162573$  & $2.9019278$  \\
9 & $-7.477946240$ & $0.302062836$  & $2.9025948$  \\
10 & $-7.477975286$ & $0.302006041$  & $2.9033303$  \\
&&& \\
Yan \emph{et al.} \cite{YanDra:95a,Yanetal:96a}& $-7.47806032310(31)$		&$0.301842809(15)$ 	&$2.905922(50)$ \\
\end{tabular}
\end{indented}
\end{table}

\begin{table}[!ht] 
\caption{\label{tab:2P_SP+T}Energies, $S_{\textrm{sms}}$ and hyperfine parameters obtained using the PCFI method for the $^2P^o$ state.}
\begin{indented}
\item[]\begin{tabular}{c | l l l}
\hline
\multicolumn{4}{c}{CAS-CP-DPCFI $\{\underline{1s},\underline{2p}\}$ } \\
\hline
$n_{max}$ & Energy (a.u.) & $S_{\textrm{sms}}$ & $a_{\textrm{orb}}$ \\
\hline
HF	&$-7.365069658$	&$-0.041898309$	&$0.0585716$ \\
4 & $-7.408715289$ & $0.249999650$ & $0.0611674$ \\
5 & $-7.409489337$ & $0.248269629$ & $0.0626329$ \\
6 & $-7.409782422$ & $0.247618411$ & $0.0627899$ \\
7 & $-7.409921067$ & $0.247249783$ & $0.0628740$ \\
8 & $-7.409996500$ & $0.247032302$ & $0.0629599$ \\
9 & $-7.410041507$ & $0.246907864$ & $0.0629917$ \\
10 & $-7.410071044$ & $0.246830104$ & $0.0630317$ \\
&&& \\
Yan \emph{et al.} \cite{YanDra:95a}& $-7.4101565218(13)$&$0.24673781(71)$ &$$ \\
\hline
$n_{max}$ & $a_{\textrm{dip}}$ & $a_{\textrm{cont}}$ & $b_{\textrm{quad}}$\\
\hline
HF	&$-0.0117143$	&$0.0000000$ &$-0.0234287$\\
4 & $-0.0130095$ & $-0.2078035$ & $-0.0220013$ \\
5 & $-0.0133191$ & $-0.2122187$ & $-0.0225979$ \\
6 & $-0.0133551$ & $-0.2122960$ & $-0.0226542$ \\
7 & $-0.0133691$ & $-0.2128985$ & $-0.0226978$ \\
8 & $-0.0133940$ & $-0.2130121$ & $-0.0227275$ \\
9 & $-0.0133960$ & $-0.2133731$ & $-0.0227428$ \\
10 & $-0.0134106$ & $-0.2134760$ & $-0.0227578$ \\
&&& \\
Yan \emph{et al.} \cite{Yanetal:96a}&$$ &$-0.214860(50)$&$$\\
\end{tabular}
\end{indented}
\end{table}

\Tref{tab:2P-2S_SP+T} presents the transitions energies, line strengths and weighted oscillator strengths  obtained using the CAS-CP-DPCFI method. This table illustrates the slow convergence of the oscillator strength in both gauges. It leads to a remaining gap between the length and the velocity gauges for the $n=10$ results. We suspect two effects playing against a fast convergence: -i) the use of frozen spectroscopic orbitals fixed to the Hartree-Fock solution, -ii)
the independent optimization of the PCF orbital sets that forbids the coupling between the PCFs subspaces to capture the indirect effects in the orbital optimization. 
The unconstrained solution could be obtained by solving a general non-orthogonal MCHF problem mixing the different PCFs for the orbital SCF optimization, and substituting, at the end of each orbital self-consistent-field step, the conventional CI by a DPCFI (ie. non-orthogonal CI) calculation to get the desired deconstrained eigenvector.

\begin{table}[!ht]
\caption{\label{tab:2P-2S_SP+T}Line strength and oscillator strength for the $^2P^o\;-^2S$ resonance line.}
\begin{indented}
\item[]\begin{tabular}{ l | l | l l | l l }
\hline
\multicolumn{6}{c}{CAS-CP-DPCFI $\{\underline{1s},\underline{2l}\}$} \\ \hline 
$n_{max}$ & $\Delta \textrm{E}$ (cm$^{-1}$)  & $S_{l}$ & $S_{v}$ & $gf_{l}$ & $gf_{v}$ \\
\hline
4 & $14930.92$ & $33.4862$ & $32.7741$ & $1.51872$ & $1.48642$  \\
5 & $14907.31$ & $33.1284$ & $32.9457$ & $1.50011$ & $1.49184$  \\
6 & $14902.69$ & $33.0804$ & $32.9798$ & $1.49748$ & $1.49292$  \\
7 & $14901.97$ & $33.0639$ & $32.9835$ & $1.49666$ & $1.49302$  \\
8 & $14901.97$ & $33.0536$ & $32.9761$ & $1.49619$ & $1.49268$  \\
9 & $14902.20$ & $33.0465$ & $32.9707$ & $1.49589$ & $1.49246$  \\
10& $14902.09$ & $33.0414$ & $32.9692$ & $1.49565$ & $1.49238$ \\
&&&&& \\
CAS (n=10) & $14902.23$ & $33.0038$ & $32.9973$ & $1.49396$ & $1.49367$\\
CAS (n=10) $\{\underline{1s},\underline{2l}\}$ & $14902.48$ & $33.0075$ & $32.9885$ & $1.49415$ & $1.49329$\\
Yan \emph{et al.} \cite{Yanetal:98a}  & $14903.16176(29)$ & $33.00066933$ & $33.00081733$ & $1.4939139$ & $1.4939206$ \\
\end{tabular}
\end{indented}
\end{table}


\section{Partial deconstraining schemes} 

\begin{figure}[!ht]
\begin{center}
     \begin{minipage}[c]{.46\linewidth}
\include{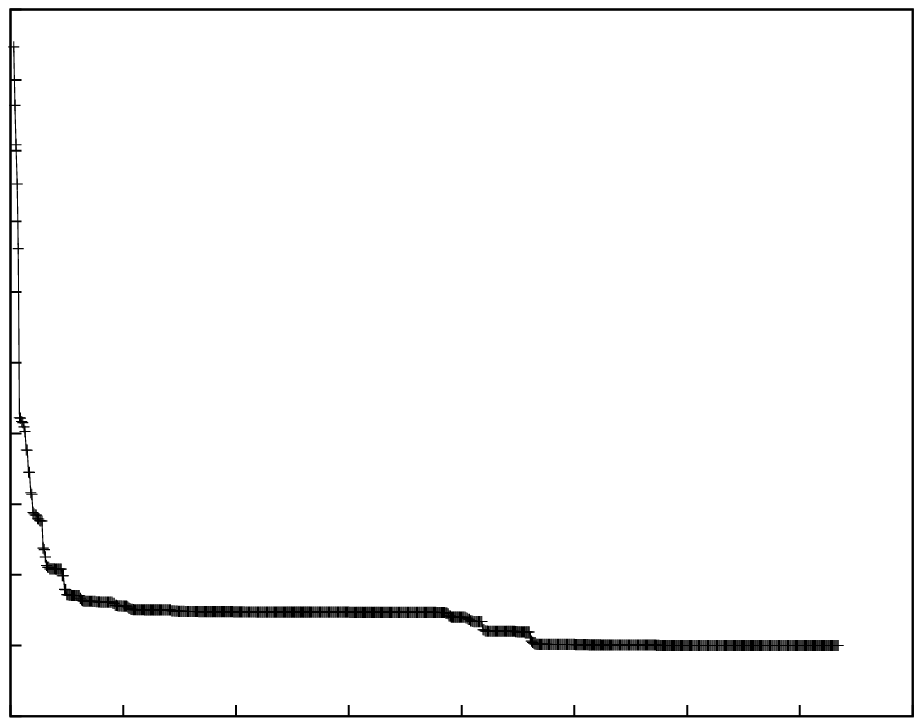} 
   \end{minipage} \hfill
   \begin{minipage}[c]{.46\linewidth}
\include{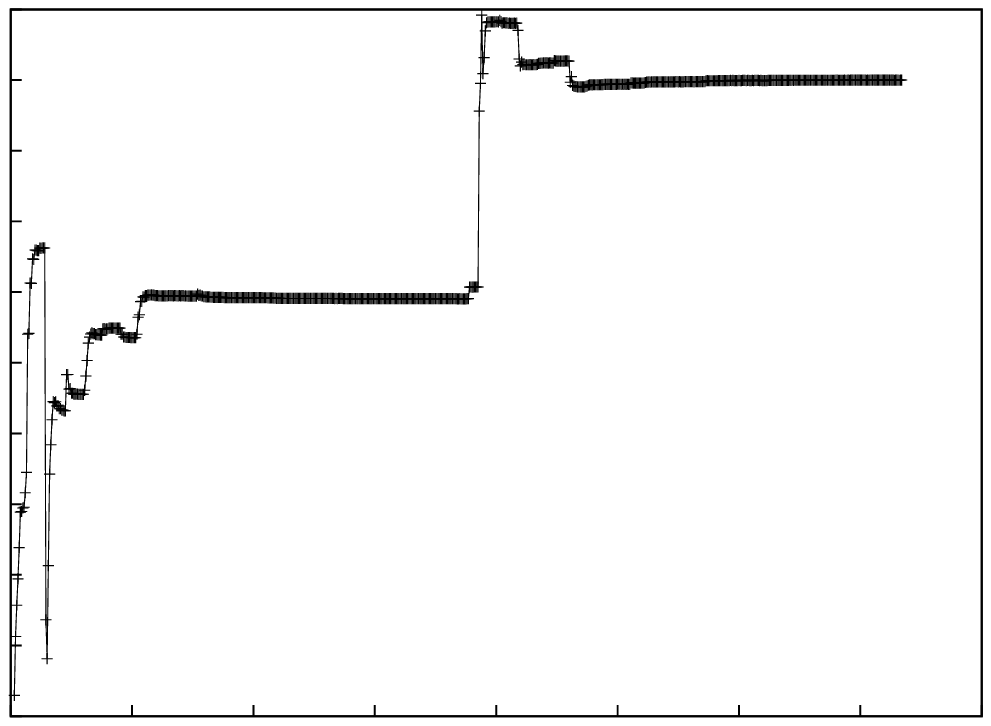}
   \end{minipage}
\include{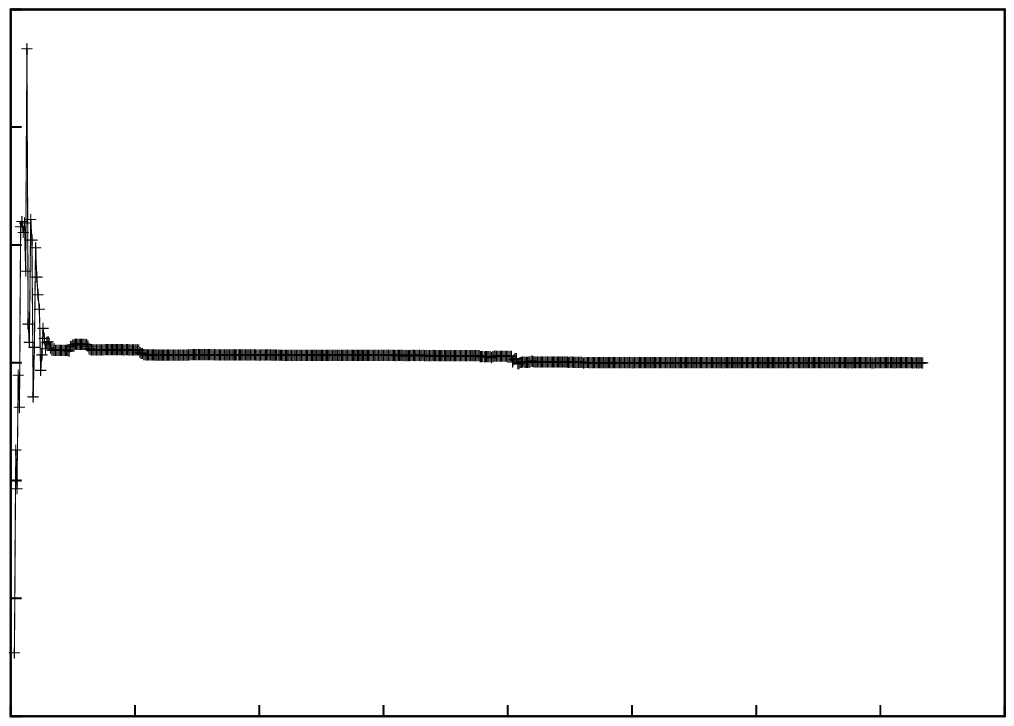}
\end{center}
\caption{Progressive deconstraint of the many-electron wave function. The three curves illustrate the relative difference between the CI calculation and the partially deconstrained problem for respectively the  total energy, the specific mass shift parameter and the hyperfine contact paremeter for the ground state $1s^2 2s \; ^2S $ of neutral lithium.}
\label{fig:progress_decon}
\end{figure}

It is worthwhile to investigate how indirect effects are distributed over CSFs. To answer this question, we  focus our analysis on the mixing coefficient constraint itself by eliminating the impact of the orbital optimization. We are therefore 
 diagonalizing the Hamiltonian operator within each PCF space, adopting  the same orthonormal orbital common set for all PCFs. At each step we promote one selected CSF, included in one PCF, at the same level than the reference CSF and then we solve the associated eigenvalue problem for finding the total energy and the two other properties.
In this way, we progressively remove all constraints in the expansion coefficients, going from a low-dimension PCFI to a larger CSF-CI calculation.
 Our selection rule for choosing at each step the promoted CSF is somewhat arbitrary since we adopt the order of the configuration list produced by {\it lsgen} program~\cite{StuFro:93a}. In the present case, the hierarchy follows i) S from the valence, ii) S from the core, iii) D from core-valence and iv) D from the core. \Fref{fig:progress_decon} illustrates the evolution, for each property, of the relative difference between the  value calculated for a given matrix size, associated to a partially deconstrained many-electron wave function, and the corresponding CSF-CI result.
\Fref{fig:progress_decon} shows not only that the operators are affected differently by the constraint, but also that the way in which the constraint effect is distributed over the CSFs directly depends on the selected operator itself. One may note that, accordingly to the variational principle, the  total energy is monotonically decreasing until it reaches the CI value, but the evolution is much more erratic for the two other properties. The presence of many plateaus in the graphs of \fref{fig:progress_decon} suggests that some efficient partial deconstraint scheme might be found. For the contact term, the latter is accidentally revealed,  as illustrated by the third graph of \fref{fig:progress_decon}, due to the fact that the most important contributions arise from the the single excitations from the core that appear first in the generated list. 
For the  $S_{\mathrm{SMS}}$  parameter, the jump occurring around $400$ corresponds to the presence of $2s 2p^2 \; ^2S$ in the expansion of $\Lambda_{1s-1s1s}$. This excitation is introduced quite late in the sequence since it is generated by {\it lsgen} in the fourth category (double excitations from the core).  \\

 By selecting and deconstraining the CSFs which constitute the dominant contributors to the constraint effect, it should be possible to optimize the CSF space partition that produces the more efficient PCF basis. We will show that full deconstraining is not always necessary and that some deconstraining schemes might be better than others. 
The key point in this analysis is to find in the CF space, for a given property, the best candidates for a possible promotion in the CSF transfer~\eref{eq:transfer}. 
As illustrated by figure~\ref{fig:progress_decon}, it is hard to predict the behavior of a given property with respect to a particular scheme of deconstraint, except for the total energy, thanks to the Hylleraas-Undheim theorem~\cite{BraJoa:83a}. 
It is here interesting to remember what we learn from the time-independent perturbation theory. Defining the $\vert n^{(0)} \rangle$ states as the eigenstates of the zeroth-order Hamiltonian $H^{(0)}$, and introducing a perturbation $V$ such as  $H= H^{(0)} + \lambda V$,  the eigenfunction $\vert n \rangle$ of $H$ can be expressed as the following $\lambda$-expansion
\begin{eqnarray}
\label{wfn_pert}
\vert n \rangle & = &  \vert n^{(0)} \rangle  + \lambda \sum_{k \neq n} \vert k^{(0)} \rangle  \frac{V_{kn}}{E^{(0)}_n - E^{(0)}_k} \\
& + & \lambda^2 \left(\sum_{k \neq n} 
\sum_{l \neq n} \frac{ \vert k^{(0)} \rangle V_{kl} V_{ln}}{(E^{(0)}_n - E^{(0)}_k)(E^{(0)}_n - E^{(0)}_l)}
- \sum_{k \neq n}  \frac{ \vert k^{(0)} \rangle V_{nn} V_{kn}}{(E^{(0)}_n - E^{(0)}_k)^2}  \right) + \ldots , \nonumber
\end{eqnarray}
where the $V_{ij} = \langle i^{(0)} \vert V \vert j^{(0)} \rangle$ matrix elements are calculated using the zero-order basis~\cite{Sak:85a}. From this expression, it is easy to derive the $\lambda$-expansion of the expectation value of any operator~$X$
\begin{eqnarray}
\langle  n \vert X \vert n \rangle & = &   \langle  n^{(0)}  \vert X \vert n^{(0)} \rangle
   + \lambda \left( \sum_{k \neq n}      \frac{X_{nk} V_{kn}}{E^{(0)}_n - E^{(0)}_k}
                 +  \sum_{k \neq n}      \frac{X_{kn} V_{kn}}{E^{(0)}_n - E^{(0)}_k}   \right) 
    \nonumber \\
&& + \lambda^2 
 \left(  \sum_{k \neq n} \sum_{k' \neq n}   \frac{X_{k k'} V_{kn}V_{k'n}}{(E^{(0)}_n - E^{(0)}_k)(E^{(0)}_n - E^{(0)}_{k'})} \right.
 \\
& & \left. + \sum_{k \neq n}
\sum_{l \neq n} \frac{  X_{nk} V_{kl} V_{ln}}{(E^{(0)}_n - E^{(0)}_k)(E^{(0)}_n - E^{(0)}_l)}
- \sum_{k \neq n}    \frac{  X_{nk}  V_{nn} V_{kn}}{(E^{(0)}_n - E^{(0)}_k)^2}  \right. \nonumber \\
& +  & \left. \sum_{k \neq n} 
\sum_{l \neq n} \frac{  X_{kn} V_{kl} V_{ln}}{(E^{(0)}_n - E^{(0)}_k)(E^{(0)}_n - E^{(0)}_l)}
- \sum_{k \neq n}    \frac{  X_{kn}  V_{nn} V_{kn}}{(E^{(0)}_n - E^{(0)}_k)^2}  \right)  + \ldots \nonumber
\end{eqnarray}
The first term is the reference value. The ${\cal O} (\lambda)$ terms are called the ``crossed second-order'' (cso) corrections to the zeroth-order expectation value $\langle X \rangle_n$ since they involve both the Hamiltonian and the operator $X$  in the coupling matrix elements. These cso contributions have been used successfully in the parametric method to analyze isotope shifts~\cite{Bau:69a}, hyperfine structures~\cite{Jud:63a} and field shifts~\cite{Auf:82a}. 
From this expression, one can realize that it would be interesting for an efficient ab initio approach, to partially deconstrain the PCFs according to their contribution to the expectation value. The CSFs that are the best candidates for a transfer promotion~\eref{eq:transfer} are the ones that are coupled to the reference by both the Hamiltonian {\it and} the relevant operator ($X_{nk} \neq 0 ; V_{nk} \neq 0)$ but these are not the only ones.  
For the SMS parameter of the lithium ground state, the HF value is strictly zero. The non zero value 
is made of correlation components only 
\begin{equation}
\label{eq:sms_cont}
S_{\mathrm{sms}} = \sum_{ij} c_i c_j \langle \Phi_i \vert X_{\mathrm{sms}} \vert \Phi_j \rangle = 0.301~450~504
\end{equation}
and a close ranking analysis of the  SMS matrix shows that the first ten contributions to~\eref{eq:sms_cont} reported in~\Tref{tab:2S_cso} account for 95\% of the total contributions.

\begin{figure}
\begin{center}
\includegraphics[scale=0.42]{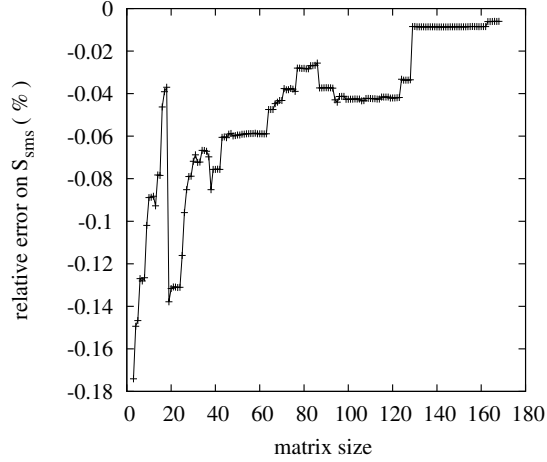}
\caption{Li $1s^2 2s \; ^2S$. Partial deconstraint according to the SMS contributions sorting.}
\label{fig:Li_cso}
\end{center}
\end{figure}

\normalsize
\begin{table}[!ht] 
\caption{\label{tab:2S_cso}The first ten more important contributions to \eref{eq:sms_cont} for  the lithium ground state $S_{\textrm{sms}}$ parameter.}
\begin{indented}
\item[]\begin{tabular}{rllrc}
\hline
\hline
   &  $\Phi_i$  &  $\Phi'_j$ & $ 2 \langle \Phi_i \vert X_{\mathrm{sms}} \vert \Phi_j \rangle$ \\
\hline
1.  & $ 1s^2  2s              $  &  $ 2s  2p^2 (^1S)       $ & $ 0.31761489 $ \\
2.  & $ 2s  2p^2 (^1S)       $  &  $  2s 3s  $ & $              -0.01267429 $ \\  
3.  & $ 1s^2  2s             $  &  $  1s  2p [^1P] 4p  $ & $     0.01141852 $\\ 
4.  & $ 2s  2p^2             $  &  $  2s  3d^2 (^1S) $   & $    -0.00580568 $\\
5.  & $ 2s  3s^2             $  &  $  2s  3p^2 (^1S) $   & $    -0.00397443 $\\
6.  & $ 1s^2  2s             $  &  $  2s  2p [^3P] 3p  $ & $    -0.00342031 $\\
7.  & $ 1s^2  2s            $  &  $  2s  3p^2 (^1S) $    & $     0.00154986 $\\   
8.  & $ 1s^2 2s           $  &  $  2s  2p [^1P] 3p $ &     $    -0.00111510 $\\   
9.  & $ 2s  3p^2 (^1S)       $  &  $  2s  3d^2 (^1S) $ &   $    -0.00110056 $\\
10. & $ 2s  2p [^3P] 3p $  &  $  2s( 1)  3s( 2) $ &        $     0.00109413 $\\
11. & \ldots & \ldots & \ldots \\
&& \\
 \\
\end{tabular}
\end{indented}
\end{table}
%
%
 In a perturbation approach, the contributions appearing in lines~$( 1,3,6,7,8)$ of~\tref{tab:2S_cso} would be cso contributions of ${\cal O} (\lambda)$. The other five higher-order corrections are rather important too  and would result from contributions of the type
($X_{k k'} V_{kn}V_{k'n}$). If we use, for the progressive PCF deconstraint, the list of the CSFs sorted according to their contribution to the $S_{\textrm{sms}}$ parameter  we get a very efficient deconstraining scheme for this property. \Fref{fig:Li_cso} illustrates indeed that the transfer of the $\simeq~140$ first CSFs produces a rather good result, to put in contrast with \Fref{fig:progress_decon}.

For the total energy, the most natural deconstraining scheme is based on a the weight criteria. The importance of a CSF is indeed mainly determined by the magnitude of its mixing coefficient that can be estimated from the first-order correction to the wave function (the  ${\cal O} (\lambda)$ term in \eref{wfn_pert}), i.e.
\begin{equation}
c_k \approx   \frac{H_{kn}}{E^{(0)}_n - E^{(0)}_k} 
\end{equation}
To illustrate the gain of partial deconstraint, we focus on Be~$1s^2 2s 2p \; ^1P^o$ considered in
\sref{subsec:Be_DPCFI}. Starting from the constrained representation corresponding to the PCFI solution, we deconstrain, 
in each of the three MR-PCF wave function~\eref{eq:MR-PCF}, all CSFs having a mixing coefficient larger than a given threshold, ie.~$\vert c_j^\Lambda \vert \geq \epsilon$. Repeating this operation for a decreasing threshold $\epsilon$, we progressively move from the constrained PCFI solution ($\epsilon = 1$) to the deconstrained DPCFI one ($\epsilon = 0$).
Using $\epsilon = 1.~10^{-4}$, we reduce the computer time by a factor of $10-20$ relatively to the DPCFI calculations.
For the $n=10$ active set, the size of the eigenvalue problem is respectively $9$ (PCFI), $788$ (partially deconstrained) and $21~632$ (DPCI) wave functions.  The values for the total energy, the $S_{\mathrm{SMS}}$ and hyperfine parameters are compared with the HF, MR, $(n=10)$ SD-MR-MCHF,  PCFI and DPCFI results in \Tref{tab:pPCFI}.
\begin{table}[!ht]
\caption{\label{tab:pPCFI}Comparison between constrained PCFI, partially deconstrained ($\epsilon = 1.~10^{-4}$) and DPCFI results for Be~$1s^2 2s2p \; ^1P^o$.}
\begin{indented}
\item[]\begin{tabular}{ c  r  c c  l l }
\hline
& $n \leq $ & $E~(E_{\mathrm{h}})$ & $S_{\mathrm{SMS}}$ (a$_0^{-2}$) &$a_{orb}$ (a$_0^{-3}$) & $b_{quad}$ (a$_0^{-3}$)\\
\hline
&&&&& \\
HF &  2        &  $-14.394735414$ & $0.004136928$ & $0.0917535$ & $-0.0367014$\\
MR &  3        &  $-14.421696066$ &$-0.000135774$ & $0.1563725$ & $-0.0590602$\\
SD-MR-MCHF & 10 & $-14.473005606$ & $0.434722502$ & $0.1846345$ & $-0.0672910$ \\
&&&&& \\
 &7 	&  $-14.472842258$ & $0.434834517$ & $0.1837046$ & $-0.0667457$ \\
partially  & 8 	&  $-14.473018319$ & $0.434557950$ & $0.1838054$ & $-0.0656012$ \\
deconstrained SD-MR-PCFI   & 9 	&  $-14.473117735$ & $0.434469982$ & $0.1839501$ & $-0.0659210$ \\
& 10 	&  $-14.473176238$ & $0.434340871$ & $0.1839869$ & $-0.0660065$ \\
&&&&& \\
SD-MR-PCFI       & 10 & $-14.473110086$ & $0.439248058$ & $0.1769675$ & $-0.0643341$ \\
SD-MR-DPCFI      & 10 & $-14.473185865$ & $0.434103182$ & $0.1844820$ & $-0.0675441$ \\
&&&&& \\
\end{tabular}
\end{indented}
\end{table}
The $(n=10)$ PCFI result is already better than the $(n=10)$ SD-MR-MCHF value for the total energy. 
 Taken the DPCFI results as the reference for the other parameters, decontraction becomes important as discussed in \sref{constraint} but partial deconstraint with $\epsilon = 1.~10^{-4}$ allows to produce quite accurate results, recapturing 96~\% ($S_{\mathrm{SMS}}$), 93~\% ($a_{orb}$) and 52~\% ($b_{quad}$) of the way to the DPCFI values. The fact that only 52 \% of $b_{quad}$ is recaptured once again illustrates the weak coupling of this quantity to the energy.


\section{Conclusions and perspectives} 

We extend the multi-configuration Hartree-Fock method by partitioning the correlation function space into several subspaces to get a better description of the dynamical correlation.
 For each of the subspaces, we optimize a partition correlation function to correct the multireference wave function. The atomic state is described by the eigenstate of a low-dimensional  PCFI eigenvalue problem, regrouping the different PCFs with the multireference function. 
 By relaxing the orthonormality restriction on the orbital basis between the different subspaces, one obtains several mutually non-orthogonal orbital basis sets that are better adapted to the short range nature of the dynamical correlation. 

Our original work~\cite{Veretal:2010a}, targeting the ground state of beryllium, leads to some unexpected complication when the PCFI eigenvector is used for the calculation of spectroscopic properties other than the total energy. We identified the source of the problem as the loss of variational freedom in solving the interaction eigenvalue problem that determines the PCF weights. 
We evaluated the differences in terms of convergence between properties calculated, on the one hand with a single orthonormal basis spanning an MCHF wave function in the traditional approach, and on the other hand with several tailored orbital basis sets optimized for different PCFs and underlying a (D)PCFI wave function.
 In addition to energy we consider specific mass shift, hyperfine structure parameters and transition data. 
We demonstrate that by (partially) deconstraining the mixing coefficients within each PCF subspace, one converges to the correct limits and keeps the tremendous advantage of improved convergence rates that comes from the use of several orbital sets. Reducing ultimately each PCF to a single CSF with its own orbital basis leads to a non-orthogonal configuration interaction approach. 

We found that partial deconstraining schemes are very attractive. For the energy indeed, the PCFI approach is shown to be highly efficient by producing accurate eigenvalues. Partial deconstraining schemes based on the CSF weight criteria appear efficient and natural for the Hamiltonian. 
For other properties, we demonstrated that a deconstraint scheme based on the ranking of the contributions to the relevant observable becomes much more efficient, as expected from a perturbation analysis. The PCFI variational method, assisted by such a perturbation analysis to define the CSFs that should be targeted in the deconstraining strategy, is extremely promising in view of the present results. The interest of the method is the fast convergence of the desired observable combined with a large reduction factor in the size of the interaction problem (and the expected CPU time gain factor that goes with it), added to the possibility of parallelizing the PCF-optimization processes before regrouping the various subspaces with the reference function to build the atomic state.

The (D)PCFI method is also highly flexible. Taking advantage of the possibility of tailoring the correlation subspace partitions, we built a specific partition correlation function for capturing core-polarization effects. The addition of this PCF tremendously improves the convergence of all the hyperfine parameters. This  approach provides values of the total energy, specific mass shift and hyperfine parameters with an acceptable accuracy for rather small ($n=6$) orbital active sets.
 These calculations demonstrate that it becomes easy to gain freedom for capturing effects weakly connected to energy, improving the convergence of atomic properties other than the energy. 

For the three- and four-electron systems considered in the present work, the (D)PCFI method might not appear as really profitable since rather complete MCHF calculations with a single orthonomormal orbital basis can be still performed. For larger systems however, as explained in the introduction, the scaling wall becomes the  real barrier and there is no hope to capture the dynamic correlation between electrons in all the different shells.  The breaking down of the very large calculations based on a common orbital set into a series of smaller parallel calculations  will then become really relevant and opens new perspectives. With this respect, five- and six-electron systems that are beyond the current limits of the Hylleraas method are the first targets in our research list. 

The proposed (D)PCFI method can be applied to any multiconfiguration approach. We will soon start to implement it in the fully relativistic codes~\cite{Jonetal:2012a} but the present calculations in the non-relativistic framework~\cite{Froetal:2007a} remain the ideal benchmarks for developing new computational strategies. Another natural line of research is to investigate how the remaining constraints on radial orbitals could be fully relaxed. This long term project could lead to  a more general non-orthogonal MCHF approach corresponding to the unconstrained SCF-DPCFI approach.

In quantum chemistry,  a method for a one-component relativistic treatment of molecular systems using a multiconfigurational approach (CASSCF) with dynamical correlation treated with second order perturbation theory (CASPT2) has been proposed by Malmqvist \etal~\cite{Maletal:2002a}.
In their work, the Hamiltonian matrix is obtained by an extension of the restricted active space state interaction (RASSI) method. They also use non-orthogonal orbitals and open some interesting perspectives that we might investigate for future work, combining 
RASSI and the present PCFI approach.

\ack

S. Verdebout had a F.R.I.A. fellowship from the F.R.S.-FNRS Fund for Scientific Research.
M. Godefroid thanks the Communaut\'e fran\c{c}aise of Belgium (Action de Recherche Concert\'ee)
and the Belgian National Fund for Scientific Research (FRFC/IISN Convention) for financial support. Financial support by the Swedish Research Council is gratefully acknowledged.

\clearpage

\appendix

\section*{Appendix 1: The PCF building rules for a unit PCFs overlap matrix.}

Let us assume that the MR function is built on the reference orthonormal subset~$\{ \phi_k \}$~:
\begin{itemize} 
\item $ | \Psi^{\textsc{mr}} \rangle  \; ; \{\phi_1,...,\phi_n \} $.
\end{itemize}
and consider two different PCFs, $\Lambda_1$ et $\Lambda_2$, built on the MR orbitals and on independent one-electron orbital sets~$\{ \theta_k \}$ and ~$\{ \zeta_k \}$,

\begin{itemize}
\item $ | \Lambda_1 \rangle  $ using an  orthonormal orbital basis 
$ \{\phi_1,...,\phi_n \;  ; \; \theta_1,...,\theta_{m_1} \} $,

\item  $ | \Lambda_2 \rangle $ using {\it another}  orthonormal orbital basis 
$\{ \phi_1,...,\phi_n \; ; \;  \zeta_1,...,\zeta_{m_2} \} $ ,
\end{itemize}
both built from excitations of the (multi-)reference function $\Psi^{\textsc{mr}}$.
In our approach, the reference orbitals {\it sub}sets  $\{ \phi \}$ of both PCFs $\Lambda_1$ and $\Lambda_2$ are  necessarily identical, being taken from the optimization of the reference function~$(\Psi^{\textsc{mr}})$. \\

In second quantization, one can write the orbital orthonormality within the three subsets $\{ \phi \}$, $\{ \theta \}$ and $\{ \zeta  \}$ as follows\[
 \{ \phi_a, \phi_b \} =  \{ \phi^\dagger_a, \phi^\dagger_b \} = 0 \hspace*{1cm} \mbox{and}   \hspace*{1cm} \{\phi_a, \phi^\dagger_b \} = \delta_{a,b}
  \]
\[ 
\{ \theta_a, \theta_b \} =  \{ \theta^\dagger_a, \theta^\dagger_b \} = 0 \hspace*{1cm} \mbox{and}   \hspace*{1cm}  \{ \theta_a, \theta^\dagger_b \}  = \delta_{a,b}
\]
\[
 \{ \zeta_a, \zeta_b \} = \{ \zeta^\dagger_a, \zeta^\dagger_b \} = 0 \hspace*{1cm} \mbox{and}   \hspace*{1cm}  \{ \zeta_a, \zeta^\dagger_b \} = \delta_{a,b} \; .
\]
The orbital orthogonality being imposed {\it within} each PCF, we also have
\[
 \{ \phi_a, \theta^\dagger_b \} = \{ \phi_a, \zeta^\dagger_b \} = 0  \; .
\]
On the other hand, the orthogonality between $\{ \theta \} $ and $\{ \zeta \}$ orbital subsets is lost :
\[
\{ \theta_a, \zeta_b \} = \{ \theta^\dagger_a, \zeta^\dagger_b \} = 0 \hspace*{1cm} \mbox{with}   \hspace*{1cm}
\{ \theta_a, \zeta^\dagger_b \} = S^{\theta \zeta}_{a,b} \; .
\]
\vspace*{0.2cm}

An arbitrary PCF can be written as a linear combination of CSFs and consequently, of Slater derminants (Sdets). We will show that the overlap between any pair of Sdets appearing in the overlap matrix element involving two different PCFs is zero. A corollary is that the overlap between PCFs is inevitably zero, ie. $ \langle \Lambda_i \vert \Lambda_j \rangle  = \delta_{ij}$.


\noindent Let 
\begin{itemize}
\item $ \langle 0 | \phi_1 ... \phi_n ... \theta_1 ... \theta_n  $ \; ,  an arbitrary Sdet belonging to PCF $ \langle \Lambda_1 \vert$, 
\item $ \zeta^\dagger_n ...  \zeta^\dagger_1 ... \phi^\dagger_n ... \phi^\dagger_1 | 0 \rangle $ \; ,  another Sdet belonging to the other PCF $\vert \Lambda_2 \rangle$.
\end{itemize}

\noindent 
Let us assume that the PCFs are built according  the following rules:
\begin{enumerate}
\item 
the excitation prototypes appearing in different PCFs have necessarily different {\it occupation} numbers for the reference orbitals~$\{ \phi_k \}$,
\item 
there is no CSF redundancy ($\Lambda_i \cap \Lambda_j = \varnothing$),
\item 
the orthogonality constraints between the variational orbitals, $\{ \theta_k \}$ in $\Lambda_1$, and $\{ \zeta_k \}$ in $\Lambda_2$,  and the (frozen) reference ones $\{ \phi_k \}$ are imposed as usual, through the Lagrange multipliers of the MCHF equations.
\end{enumerate}
 
By respecting these ``{\it building rules}'' for PCFs constructed from excitations of orbitals $\{ \phi_k \} $  spanning $ \Psi^{\textsc{mr}}$, we are assured that Sdets belonging to different PCFs have different occupation numbers for the reference orbitals  $\{ \phi_k \}$.
This implies that {\it at least} one of the creation operators associated with a reference spin-orbital appears {\it only} in one of the two Sdets, and {\it not} in the other one. This creation operator is then free to  ``sail'' in the creation/annihilation operator sea, with a possible phase factor as the only price to pay. Moving it to the left/right hand-side  until it acts on the bra/ket vacuum state, we ultimately get a null result ({\it qed})
\\

\noindent As an illustration, if the  $ \phi_2$ spin-orbital does not appear in the bra, we have~:
\[
\langle 0 | \phi_1 \phi_3... \phi_n ... \theta_1 ... \theta_n \zeta^\dagger_n
...\zeta^\dagger_1 ... \phi^\dagger_n ... \phi^\dagger_2 \phi^\dagger_1 | 0 \rangle
\]
\[  =
(-1)^{N_p}
 \langle 0 |\phi^\dagger_2 \phi_1 \phi_3... \phi_n ... \theta_1
... \theta_n \zeta^\dagger_n ...\zeta^\dagger_1 ... \phi^\dagger_n ...  \phi^\dagger_1 | 0 \rangle = 0 
\]
where $N_p$ is the required number of permutations. \\

\noindent 
We have demonstrated that the PCFs overlap matrix is the unit matrix if we respect the above PCF building rules. There are however situations for which {\it violating these rules} might be interesting from the variational point of view. In these cases, the PCF overlap matrix is not the unity matrix anymore, forcing us to treat the PCFI interaction problem as the {\it generalized} eigenvalue problem \eref{GEP}.  \\

\noindent The non-diagonal PCF overlap matrix element $\langle \Lambda_1 \vert \Lambda_2 \rangle$ will indeed differ from zero if
\begin{itemize}
\item there exist two CSFs, belonging respectively to $\Lambda_1$ and $\Lambda_2$, that have identical occupation numbers for all the reference orbitals,
\item the reference orbitals cannot be strictly qualified as ``reference'' ones, when another subset is used for one of the PCFs ie. $\{ \phi'_k \} \neq \{ \phi_k \}$ . For example, if the $\{ \phi'_k \}$ orbitals of $\Lambda_1$ do not have the same radial functions than the $\{ \phi_k \}$  of $\Psi^{\textsc{mr}}$ and $\Lambda_2$, although having the same labels, some commutation rules are lost :
\[ \{ \phi_a ', \zeta^\dagger_b \} = S^{\phi' \zeta}_{ab} \; , \]
\item non-orthogonalities are introduced within a given PCF, for instance within $\Lambda_2$:
\[ \{ \phi_a,\zeta^\dagger_b \} = S_{ab}^{\phi\theta} \; . \]    
\end{itemize}

\clearpage

\section*{Appendix 2: The weight matrix formalism}

In the present appendix, we will show how the whole sub-matrix block of the matrix~\eref{eq:big_mat} can be evaluated by performing a {\it single} biorthonormal transformation treating simultaneouly the counter-transformation of all the elements constituting the block-basis. For this purpose, we introduce the weight matrix ${\bf D}^i$, corresponding to the deconstrained PCF $\{ \Phi^{i}_1 , \ldots,  \Phi^{i}_{h_i}, \overline{\Lambda}_i^{d}\}$. Each column of the latter is composed of the CSF expansion coefficients for each of the $(h_i + 1)$ elements
\begin{equation}\label{eq:D}
{\bf D}^i = \left(
\begin{array}{llll}
1 && 0 & 0   \\
0 &\ddots& 1 & 0   \\
0 && 0 & d_{(h_i + 1)} \\
  &\vdots&  & \vdots \\
0 && 0 & d_{dim(\overline{\Lambda}_i)} \\
0 && 0 & 0_1 \\ 
  &\vdots&\ddots  & \vdots \\
0 && 0 & 0_{dim(\mathrm{CUD}_i)} 
\end{array}
\right)\;.
\end{equation}
The number of lines of this weight matrix is given by the size of the original PCF ($dim(\overline{\Lambda}_i)$) to which we add $dim(\mathrm{CUD}_i)$ CSFs necessary for satisfying the CUD condition.
In the deconstrained limit case, the weight matrix is a rectangular unit matrix of size $([ dim(\overline{\Lambda}_i) + dim(\mathrm{CUD}_i) ] \times dim(\overline{\Lambda}_i)) $
\begin{equation}
{\bf D}^i = \left(
\begin{array}{llll}
1 & 0 &  & 0   \\
0 & 1 &  & 0   \\
  & \vdots & \ddots &\vdots\\
0 & 0 &  & 1 \\ 
0 & 0 &  & 0_1 \\ 
  & \vdots & \ddots & \vdots\\
0 & 0 &  & 0_{dim(\mathrm{CUD}_i)}   
\end{array}
\right)\;.
\end{equation}
Adopting the weight matrix, the expression of a sub-matrix block $((h_i+1)\times(h_j+1))$ involving two different deconstrained PCFs can be obtained by a simple matrix multiplication 
\begin{equation}
\langle \{ \Phi^{i}_1 , \ldots,  \Phi^{i}_{h_i}, \overline{\Lambda}_i^{d}\}  | O | \{ \Phi^{j}_1 , \ldots,  \Phi^{j}_{h_j},\overline{\Lambda}_j^{d}\} \rangle = (\tilde{{\bf D}}^{i})^t \tilde{{\bf O}}\tilde{{\bf D}}^{j} 
\end{equation}
involving the counter-transformed weight matrix 
\begin{equation}\label{eq:Dtilde}
\hspace*{-2cm}
{\bf \tilde{D}}^i = \left(
\begin{array}{llll}
\tilde{d}^i_{1,1} & \cdots & \tilde{d}^i_{1,h_i} & \tilde{d}^i_{1,(h_i+1)}   \\
\vdots &\ddots & & \vdots \\
\tilde{d}^i_{dim(\overline{\Lambda}_i),1}  & \cdots & \tilde{d}^i_{dim(\overline{\Lambda}_i),h_i} &  \tilde{d}^i_{dim(\overline{\Lambda}_i),(h_i+1)} \\
\vdots &\ddots & & \vdots \\
\tilde{d}^i_{(dim(\overline{\Lambda}_i)+dim(\mathrm{CUD}_i)),1} & \cdots & \tilde{d}^i_{(dim(\overline{\Lambda}_i)+dim(\mathrm{CUD}
_i)),h_i} & \tilde{d}^i_{(dim(\overline{\Lambda}_i)+dim(\mathrm{CUD}_i)),(h_i+1)}
\end{array}
\right)\; ,
\end{equation}
a similar one for $\tilde{{\bf D}}^{j}$,
and the matrix representation of the selected operator $\tilde{{\bf O}}$~\eref{eq:mat_operator}. The weight matrix formalism introduces the possibility of doing the simultaneous counter-transformation, associated to the biorthonormal transformation, of the expansion coefficients of each element of the deconstrained PCF. As each column of the original weight matrix~\eref{eq:D} represents the expansion coefficients of each element, i.e. $\Phi^{i}_1 , \ldots,  \Phi^{i}_{h_i}$ and $\overline{\Lambda}_i^{d}$, of a deconstrained PCF in the original CSF space, each column of the counter-transform weight matrix~\eref{eq:Dtilde} gives the representation of these elements in the new CSF space expressed in the biorthonormal orbital basis.
 Another advantage of the weight matrix formalism is that we are able to treat, without any distinction, an arbitrary degree of deconstraint. The price to pay is a simple matrix dimension adaptation. By allowing high degrees of deconstraint, we quickly get large weight matrices and the use of their sparse representation is welcome. 

For simplifying the counter-transformation process associated to the biorthonormal transformation, the {\it lscud} program originally developed for closing a CSF expansion under de-excitation ~\cite{Veretal:2010a} was adapted to provide
 the sparse structure of the  counter-transformed weight matrix ${\bf \tilde{D}}^i$ before performing the biorthonormalisation by itself. This is made possible by the use of the Warshall's algorithm~\cite{Kos:2003a} applied to the adjacency matrix mapping the single de-excitation operator for all $l$-values within the CSF basis. The obtained transitive closure matrix of the single de-excitation operator corresponds to the mask of the ${\bf \tilde{D}}^i$ matrix that determines the position of the non-zero elements after biorthonormalisation.

\clearpage

%
%

\section*{References}
\bibliographystyle{unsrt}

\end{document}